\documentclass[structabstract]{aa}
\usepackage{txfonts}
\usepackage[pdftex]{graphicx}
\usepackage{natbib}

\bibpunct{(}{)}{;}{a}{}{,}

\begin{document}

\newcommand{\be}{\begin{equation}}
\newcommand{\ee}{\end{equation}}
\newcommand{\bdm}{\begin{displaymath}}
\newcommand{\edm}{\end{displaymath}}
\newcommand{\bea}{\begin{eqnarray}}
\newcommand{\eea}{\end{eqnarray}}

\newcommand{\cf}{\textit{cf.}~}
\newcommand{\ie}{\textit{i.e.}~}

\newcommand{\lr}[1]{ \textcolor{blue}  {\texttt{\textbf{LR: #1}}} }
\newcommand{\oz}[1]{ \textcolor{red}   {\texttt{\textbf{OZ: #1}}} }

\title{EM counterparts of recoiling black holes: general relativistic 
simulations of non-Keplerian discs}

\author{
        O. Zanotti    \inst{1}
\and    L. Rezzolla      \inst{1,2}
\and    L. Del Zanna  \inst{3} 
\and    C. Palenzuela    \inst{4,1}
}

\institute{
Max-Planck-Institut f{\"u}r Gravitationsphysik, Albert Einstein Institut, Golm, Germany
\\ \email{zanotti@aei.mpg.de}
\and
Department of Physics, Louisiana State University,
   Baton Rouge, LA 70803 USA
\and
Dipartimento di Fisica e Astronomia,
Universit\`a di Firenze, Firenze, Italy
\and
Canadian Institute for Theoretical Astrophysics
(CITA), Toronto, Ontario, Canada
}

\date{}

\authorrunning{O.~Zanotti et al.}
\titlerunning{EM counterparts of recoiling black holes: general relativistic 
simulations of non-Keplerian discs}

\abstract
{}
{We investigate the dynamics of a circumbinary disc that responds to
the loss of mass and to the recoil velocity of the black hole produced
by the merger of a binary system of supermassive black
holes.} 
{We perform the first two-dimensional general
relativistic hydrodynamics simulations of \textit{extended}
non-Keplerian discs and employ a new technique to construct a ``shock
detector'', thus determining the precise location of the shocks
produced in the accreting disc by the recoiling black hole. In this
way we can study how the properties of the system, such as the spin,
mass and recoil velocity of the black hole, affect the mass accretion
rate and are imprinted on the electromagnetic emission from these
sources.} 
{
We argue that
the estimates of the bremsstrahlung luminosity computed without
properly taking into account the radiation transfer yield
cooling times that are unrealistically short.
At the same time we
show, through an approximation based on the relativistic 
isothermal evolution, that the luminosity
produced can reach a peak value above $L \simeq 10^{43} \ {\rm erg/s} $
at about $\sim 30\,{\rm d}$ after the merger of a binary with total
mass $M\simeq 10^6 M_\odot$ and persist for several days at values
which are a factor of a few smaller.
If confirmed by more
sophisticated calculations such a signal could indeed lead to an
electromagnetic counterpart of the merger of binary black-hole system.}
{}


\maketitle


\section{INTRODUCTION}
\label{Introduction}

Despite the fundamental role they play in gravitational-wave
astronomy, no undisputed observational evidence of the existence of
supermassive binary black holes (SMBBHs) systems has been found
yet. However, circumstantial evidence does exist for a number of
potential candidates. This is the case, for instance, of the radio
galaxy \textit{0402+379}, which shows a projected separation between
the two black holes of $7.3\, \rm {pc}$ and a total mass of $\sim
1.5\times 10^8M_\odot$~\citep{Rodriguez2006}. Similarly, the
ultraluminous infrared galaxy \textit{NGC6240} shows two optical
nuclei and is thought to be in an early merger
phase~\citep{Komossa2003}. Finally, potential candidates have been
suggested in few other cases where the two galaxies are more widely
separated, like in the pair \textit{IC694/NGC3690} hosting two active
nuclei as revealed by the presence of two distinct K$\alpha$ lines in
their X-ray spectra \citep{Ballo2004}.

In addition, there is a large family of even more uncertain SMBBH
candidates, that are spatially unresolved and whose ultimate nature
is, of course, a matter of strong debate. They include the class of
X-shaped radio galaxies, in which the observed changes in the
orientation of the black hole spin axis could be due to an ongoing
merger with a second black hole~\citep{Gopal2003}; or the class of
double-double radio galaxies presenting a pair of double-lobed radio
structures that could be the remnants of a SMBBH merger event
\citep{Liu2003}; or, finally, the class of sources showing
periodicities in the light curves, like in BL Lac Object
\textit{OJ287}~\citep{Komossa2006}.  Quite recently, also the quasar
\textit{SDSSJ0927} (at a redshift $z\approx 0.7$) has been identified
as a promising SMBBH candidate, with a mass for the primary black hole
of $M\approx 2 \times 10^9 M_\odot$ and a semi-major axis of $0.34\,
{\rm pc}$~\citep{Dotti2009}.

A strong motivation for studying the merger of supermassive binary
black hole systems comes from the fact that their gravitational signal
will be detected by the planned Laser Interferometric Space Antenna
(LISA), whose optimum sensitivity is placed in the range $(10^{-4}
\div 0.1) \ \rm{Hz}$. Considerable attention has therefore been
recently attracted by the possibility of detecting also the
electromagnetic (EM) counterpart of these events through the emission
coming from the circumbinary accretion disc that is expected to form
when the binary is still widely separated. Such a detection will not
only act as a confirmation of the gravitational wave (GW) detection,
but it will also provide a new tool for testing a number of
fundamental astrophysical issues~\citep{Haiman2009b}. More
specifically, it will offer the possibility of testing models of
galaxy mergers and accretion discs perturbation, probing basic aspects
of gravitational physics, and allowing for the measurement of the
Eddington ratio and for the determination of cosmological parameter
once the redshift is known \citep{Phinney2009}. In spite of the
presence of the disturbing effects due to weak-lensing
errors,~\citet{Kocsis2006} have computed the average number of quasars
in the three-dimensional LISA error volume and have shown that for
mergers with masses in the range $\sim 4\times (10^5\div 10^7)M_\odot$
at redshift $z\sim 1$, the error box may contain $1$ quasar with a
luminosity $L_B\sim (10^{10}\div 10^{11}) L_\odot$ 
~\citep[see also][]{Kocsis:2008,Kocsis:2008b}.

As a product of this increased interest, a number of studies have been
recently carried out to investigate the properties of these EM
counterparts either during the stages that precede the merger, or in
those following it. As an example, recent work has considered the
interaction between the binary and a dense gas
cloud~\citep{Armitage:2002,vanMeter:2009gu,Bode:2009mt,Farris:2009mt,Lodato2009,Chang2009} 
even though
astrophysical considerations seem to suggest that during the very
final stages of the merger the SMBBH will inspiral in a rather tenuous
intergalactic medium. At the same time, other scenarios not involving
matter have also been considered. In these cases the SMBBH is
considered to be inspiralling in vacuum but in the presence of an
external magnetic field which is anchored to the circumbinary
disc~\citep{Palenzuela:2009yr, Moesta:2009}. The extensive analysis
of~\citet{Moesta:2009}, in particular, has shown that, even though the
electromagnetic radiation in the lowest $\ell= 2$ and $m = 2$
multipole accurately reflects the gravitational one, the energy
emitted in EM waves is $13$ orders of magnitude smaller than the one
emitted in GW, thus making the direct detection of the two different
signals very unlikely.  The situation changes in the post-merger
phase. In this case, in fact, the EM counterpart is supposed to be
mainly due to the radiation from the circumbinary accretion disc, and,
because of that, it will contain an imprint of any strong dynamical
change produced on the disc by the merger event. There are indeed two
major such dynamical effects. The first one is the abrupt reduction of
the rest-mass of the binary, emitted away in GWs, which is a function
of the binary mass ratio, amounting up to $\simeq 10\%$ for equal-mass
spinning systems~\citep{Reisswig:2009vc}.  The second one is the
recoil velocity of the merged system, resulting in a {\em kick}
velocity of the resulting black hole with respect to the hosting
galaxy (see~\citet{Bekenstein1973},~\citet{Redmount:1989}) for a first
discussion of the process and~\citet{Rezzolla:2008sd} for a recent
review).  Leaving aside possible problems due to the actual value of
the kicked velocity, which in some cases could be even larger than the
escape velocity, it is clear that both events mentioned above can
significantly affect the dynamics of the circumbinary disc, mainly as
they contribute to the formation and propagation of shocks, thus
enhancing the possibility of a strong EM counterpart.

After the first smooth-particle-hydrodynamics approach to the
dynamical evolution of circumbinary discs performed
by~\citet{Artymowicz1994}, several additional numerical investigations
have been proposed in the very recent past. \citet{MacFadyen2008}, for
example, performed two dimensional hydrodynamical simulations and
studied in detail the evolution of the binary separation and of the
disc eccentricity.  By perturbing Keplerian orbits of collisionless
test particles, on the other hand,~\citet{Lippai:2008} found a clear
spiral shock pattern in the plane of the disc as a response to the
kick. By performing pseudo-Newtonian numerical simulations of
Keplerian discs~\citet{Oneill2009} have recently questioned the
contribution of the shocks to the expected bremsstrahlung
emissivity, while~\citet{Megevand2009} showed that the intensity of
bremsstrahlung luminosity is not much affected by the magnitude of the
kick velocity, provided this is less than the smallest orbital speed
of the fluid. Although they represent the first fully general
relativistic calculations of this process, the simulations
of~\citet{Megevand2009} used unrealistically small discs which were
also placed extremely close to the recoiling black hole. As a
considerable improvement over all the previous
investigations,~\citet{Corrales2009} have carried out a systematic
study of the effects of the mass-loss and recoil over a number of
$\alpha$-discs in Newtonian gravity and two-dimensions. While
confirming the existence of spiral shocks, they also provided a first
realistic estimate of the resulting enhanced luminosity, which can be
as large as few $\times 10^{43}\rm{erg/s}$ when the disc is assumed to
be extremely efficient in radiating any local increase of the
temperature.  Very interesting results have also been obtained by
\citet{Rossi2010}, who estimated the maximum disc-to-hole mass ratio
that would be stable against fragmentation due to self-gravity to be
$M_d/M\sim 6\times10^{-4}$ for a supermassive black hole with mass
$M=10^6 M_\odot$. In addition, by performing three-dimensional but
Newtonian SPH simulations of geometrically thin discs, they found that
the emitted luminosity corresponding to such small disc-to-hole mass
ratios is unlikely to make the EM counterpart visible via wide-area
sky surveys.

In this paper we present the results of two-dimensional relativistic
numerical simulations of \textit{extended} circumbinary discs in the
post-merger phase of the merger, when the disc reacts to the mass loss
of the central black hole and to the received kick velocity. By
accurately capturing the dynamics of the perturbed disc in the
relativistic regime, we investigate the dependence of the accretion
rate on the black-hole spin and on the kick velocity. At the same
time, we introduce a new technique to locate the shocks that are
potentially produced by the recoil and can therefore assess under what
conditions a spiral pattern can develop, producing a variability in
the accretion rate and, hence, in the luminosity. Our ``shock
detector'' is based on the analysis of the initial states of the
Riemann problem solved at each cell interface and can therefore
determine the location of the shock with extreme precision, thus
revealing that the previously proposed criteria for the occurrence of
the shock are often inaccurate.

To compare with the general-relativistic calculations performed
by~\citet{Megevand2009}, our initial models consider {\em small-size}
discs with an inner radius at $r\sim 40 M$ and an outer one at $r\sim
120 M$. In addition, however, we also study the dynamics of {\em
  large-size} discs with an inner radius at $r\sim 400 M$ and an outer
one at $r\sim 4700 M$. These configurations have almost Keplerian
distributions of angular momentum and are therefore closer to what is
believed to be a realistic configuration for a circumbinary
disc. Furthermore, because the mass in the discs is always much
smaller than the mass of the black hole (\ie with a mass ratio $\sim
10^{-3}$), we solve the equations of relativistic hydrodynamics in the
fixed spacetime of the final black hole.

At first sight it may appear that the use of general-relativistic
hydrodynamics is unnecessary when simulating astrophysical systems
such as the ones considered here and especially for the case of
large-size discs. Such a view, however, does not take
into account that much of the
dynamics in these EM counterparts takes place near the black-hole
horizon, where general relativistic effects are not only large but
essential for a correct physical description. Moreover, we
do not have any firm theoretical basis to exclude small-size discs
and for which the relativistic corrections are non-negligible.
Finally, even in a scenario in which gravity could be
approximated by the Newton law, we cannot exclude 
the importance of special relativistic effects.

As all of the above mentioned investigations, also our approach
suffers from the absence of a a fully consistent treatment of the
radiation transfer, thus allowing only for tentative conclusions about
the energetics involved in circumbinary accretion discs. However, we
extend to the relativistic framework the strategy reported in
\citet{Corrales2009} of performing an {\em isothermal} evolution as a
tool to extract luminosity curves more realistic than those obtained
from thermal bremsstrahlung, although it exaggerates some features of
the dynamics, such as the formation of shocks.

The paper is organized as follows. In Section~\ref{Numerical_method}
we provide the essential information about the numerical code adopted
in the simulations. Section~\ref{Initial_models} describes the
physical properties of the initial models, while
Section~\ref{Monitored_quantities} highlights the most relevant
diagnostic quantities used in the rest of the
paper. Section~\ref{Results} is devoted to the presentations of the
results, and, finally, Section~\ref{Conclusions} contains a summary of
our work.  We assume a signature $\{-,+,+,+\}$ for the space-time
metric and we will use Greek letters (running from $0$ to $3$) for
four-dimensional space-time tensor components, while Latin letters
(running from $1$ to $3$) will be employed for three-dimensional
spatial tensor components. Moreover, we set $c=G=1$ and we extend the
geometric units by setting $m_p/k_B=1$, where $m_p$ is the mass of the
proton, while $k_B$ is the Boltzmann constant. In this way the
temperature is a dimensionless quantity.


\section{Numerical methods}
\label{Numerical_method}

In the stationary spacetime of a Schwarzschild or Kerr black hole we
consider the time evolution of a perfect fluid described by the usual
energy momentum tensor
\be
T_{\mu\nu}=\rho h\,u_{\,\mu}u_{\nu}+pg_{\,\mu\nu},
\label{eq:T_matter}
\ee
where $u_\mu$ is the four velocity of the fluid, $g_{\,\mu\nu}$ is the
space-time metric tensor, $\rho$ is the rest-mass density,
$h=1+\epsilon + p/\rho$ the specific enthalpy (including rest-mass
energy contribution), $\epsilon$ the specific internal energy, $p$ the
thermal pressure, related to $\rho$ and $\epsilon$ through the usual
ideal-gas equation of state (EOS)
\be
p=\rho\epsilon(\gamma-1) \ ,
\ee
where $\gamma$ is the (constant) adiabatic ratio of the gas. We solve
the corresponding equations of general relativistic non-dissipative
hydrodynamics through the \texttt{ECHO} code~\citep{DelZanna2007}.
Because the dynamics of the EM
  emission takes place on a timescale which is of the order of the
  orbital one and because the latter is much shorter than the viscous
  timescale\footnote{We recall that in geometrically thin accretion
    discs the local viscous timescale is given by $t_{vis} \simeq
    r^2/({\tilde \alpha} H^2 \Omega)$, where ${\tilde \alpha}$ is the
    standard alpha parameter ad $H$ is the half-thickness of the
    disc.}, the use of inviscid hydrodynamics is indeed a very good
  approximation.

\texttt{ECHO} adopts a $3+1$ split of
spacetime in which the space-time metric is decomposed according to
\be
\mathrm{d}s^2 = \! -\alpha^2\mathrm{d}t^2+\gamma_{ij}\,
(\mathrm{d}x^i\!+\beta^i\mathrm{d}t)(\mathrm{d}x^j\!+\beta^j\mathrm{d}t),
\label{eq:adm}
\ee 
where $\alpha$ is the lapse function, $\beta^i$ is the shift vector,
and $\gamma_{ij}$ is the spatial metric tensor.  The
general-relativistic hydrodynamical equations are written in the
following conservative form
\be
\partial_t\vec{\mathcal{U}} + \partial_i\vec{\mathcal{F}}^i=\vec{\mathcal{S}},
\label{eq:UFS}
\ee
which is appropriate for numerical integration via standard
high-resolution shock-capturing (HRSC) methods developed for the Euler
equations.  The conservative variables and the corresponding fluxes in
the $i$ direction are respectively given by
\be
\vec{\mathcal{U}}\equiv\sqrt{\gamma}\left[\begin{array}{c}
D \\ \\ S_j \\ \\U
\end{array}\right],~~~
\vec{\mathcal{F}}^i\equiv\sqrt{\gamma}\left[\begin{array}{c}
\alpha v^i D-\beta^i D \\\\
\alpha W^i_j-\beta^i S_j \\\\
\alpha S^i-\beta^i U
\end{array}\right] ,
\label{eq:fluxes}
\ee
whereas the sources, in any stationary background metric,
can be written as
\be
\vec{\mathcal{S}} \equiv \sqrt{\gamma}\left[\begin{array}{c}
0 \\  \\
\frac{1}{2}\alpha W^{ik}\partial_j\gamma_{ik}+
S_i\partial_j\beta^i-U\partial_j\alpha \\ \\
\frac{1}{2}W^{ik}\beta^j\partial_j\gamma_{ik}+{W_i}^j\partial_j\beta^i
-S^j\partial_j\alpha
\end{array}\right],
\ee
where only purely spatial quantities are present.  We note that
$\sqrt{\gamma}\equiv \sqrt{-g}/\alpha$ is the determinant of the
spatial metric. The relation between the evolved conservative
variables $(D,S_j,U)$ and the primitive variables is given by
\bea
&&D   \equiv \rho\Gamma ,\\
&&S_i \equiv \rho h \Gamma^2 v_i, \\
&&U   \equiv \rho h \Gamma^2 - p, 
\label{eq:cons}
\eea
where $\Gamma=(1-v^2)^{-1/2}$ is the Lorentz factor of the bulk flow
with respect to the Eulerian observer associated to the $3+1$
splitting of the spacetime, and
\be
W_{ij} \equiv \rho h \Gamma^2 v_i v_j +p \gamma_{ij},\\
\label{eq:W} 
\ee
is the fully spatial component of the energy-momentum tensor.  In our
setup for two dimensional disc simulations we assume the Kerr
spacetime metric in Boyer-Lindquist coordinates (\ie only
$\beta^\phi\neq 0$), with the limiting case of Schwarzschild metric
for vanishing black-hole spins, and lay our coordinates in the
equatorial plane of the disc (\ie $\theta=\pi/2$).  The radial
numerical grid is discretised by choosing $N_r$ points from
$r_\mathrm{min}$ to $r_\mathrm{max}$, non-uniformly distributed
according to the following scheme
\bea
r_i &=& r_\mathrm{min} + a_1 \tan{(a_2 x_i)} \\
x_i &=& (\tilde{r}_i-r_\mathrm{min})/(r_\mathrm{max}-r_\mathrm{min})
\eea
where $a_1=(r_\mathrm{max}-r_\mathrm{min})/a_0$, $a_2=\arctan{a_0}$,
while $\tilde{r}_i$ are the coordinate points of the uniform grid from
$r_\mathrm{min}$ to $r_\mathrm{max}$. In practice, the free parameter
$a_0$ controls the extent to which the gridpoints of the original
uniform grid are concentrated towards $r_\mathrm{min}$, and we have
chosen $a_0=5$ in most of our simulations.  The actual value of $N_r$
depends on the size of the disc, and it varies between $N_r=600$ and
$N_r=1200$.  Outflow boundary conditions are adopted both at
$r_\mathrm{min}$ and $r_\mathrm{max}$.  The azimuthal grid extends
from $0$ to $2\pi$, with periodic boundary conditions, and
$N_\phi=200$.  All runs are performed with a Courant-Friedrichs-Lewy
coefficient ${\rm CFL}=1/2$.

The set of hydrodynamics equations is discretised in time with the
method of lines and the evolution is performed with a second-order
modified Euler scheme. A fifth-order finite-difference algorithm based
on an upwind \emph{monotonicity preserving} filter is employed for
spatial reconstruction of primitive variables, whereas a two-wave HLL
Riemann solver is used to ensure the shock-capturing properties
~\citep[see][for further details]{DelZanna2007}. 
The timestep is
generically chosen to be sufficiently small so that the second-order
truncation error in time is comparable with the fifth-order one in
space.

As a final remark we note that as customary in HRSC methods, we
introduce a tenuous and static ``atmosphere'' in the regions of the
fluid outside the initial model for the disc and follow the
prescription detailed in~\citet{Baiotti04} for its evolution. In
practice we set to zero the velocity field and reset to a pre-defined
floor value the rest-mass density of any cell whose density falls
below the chosen threshold value. Such a threshold is set to be $8$
orders of magnitude below the maximum rest-mass density and we have
checked that essentially identical results are obtained when changing
this value of one or more orders of magnitude. 


\section{Initial models}
\label{Initial_models}

As initial models we adopt stationary and axisymmetric configurations
that are consistent solutions of the relativistic Euler equations and
describe a fluid in sub-Keplerian rotation around a Kerr black hole of
prescribed mass and spin~\citep{Abramowicz78}. The resulting discs are
geometrically thick and they can either have a constant or, more
generally, a non-constant radial distribution of the specific angular
momentum $\ell$.  In our simulations we have considered both
\textit{``small-size''} models, for which we adopt a constant
distribution of $\ell$, and \textit{``large-size'}' models, with a
distribution of $\ell$ that, on the equatorial plane, obeys a power
law
\begin{equation}
\label{power_law}
\ell (r, \theta = \pi/2) = {\cal S} r^q ,
\end{equation}
where ${\cal S}$ is chosen to be positive, thus providing a disc
rotation that is prograde with respect to the black hole rotation.  A
detailed description of the equilibrium models for non-constant
specific angular momentum discs can be found~\citet{Daigne04}. In
particular, we have chosen a value of ${\cal S}$ such that the
resulting thick discs possess two well defined Keplerian points,
namely the ``cusp'' (which is where matter can accrete onto the black
hole) and the ``centre'' (which is where the pressure has zero
gradient); \cf Table~\ref{tab1}.

When the exponent $q$ in~(\ref{power_law}) is chosen close to $1/2$,
the rotation law tends to the Keplerian one, and the disc flattens
towards the equatorial plane. In these circumstances the vertical
structure of the disc can be essentially neglected and two-dimensional
simulations are therefore indicative of the full three-dimensional
dynamics. 
It is worth mentioning that discs with a rotation law given
by~(\ref{power_law}) have been the subject of a long-standing debate
about whether they are subject to the so called ``runaway
instability''~\citep{Abramowicz83}, which would lead to an
exponentially rapid accretion onto the black
hole~\citep{Font02a,Font02b,Zanotti03,Daigne04,Zanotti05,Montero07}. Because
the onset and development of this instability depends on the response
of the torus to the increased mass of the black hole, simulating this
instability accurately requires also the evolution of the Einstein
equations. Recent calculations of this type have been performed
by~\citet{Montero09} and reveal that the tori are indeed stable
irrespective of the angular momentum distribution, thus excluding any
role of the runaway instability in the dynamics of the discs simulated
here. However, as we will further comment in
Sect.~\ref{Shock_propagation}, other non-axisymmetric instabilities
are possible and have been indeed found.

Table~\ref{tab1} reports the main properties of the models chosen,
where the naming convention used allows to easily distinguish the
small-size models (\texttt{S*}) from the large-size ones
(\texttt{L*}), and where the number \texttt{*} in the name refers to
the spin of the black hole, thus ranging between $0.00$ and $0.99$.
As already commented in the Introduction, while the small-size models
are particularly suitable for investigating any effect of the black
hole spin, the large-size models are those that are (astro)physically
more realistic. The inner radius of these large-size models is
typically of a few hundreds of gravitational radii and represents the
size of the cavity produced by the torque of the SMBBH as
estimated from the expression deduced from 
Table~1 of ~\citet{Milosavljevic05}
\begin{equation}
r_{\rm cavity}\simeq \left(\frac{117}{\alpha_{-1}^{0.34}}\right)
\left(\frac{\eta_{-1}}{\dot{M}_{\rm Edd}}\right)^{0.24}
\left(\frac{M}{10^6M_\odot}\right)^{0.08}\!\!\! [4q/(1+q)^2]^{0.42},
\end{equation}
where $\alpha_{-1} \equiv \alpha/0.1$, $\eta_{-1} \equiv \eta/0.1$
(${\tilde \alpha}$ and ${\tilde \beta}$ being the
effective 
${\tilde \alpha}$-parameter of thin
accretion discs and the radiative efficiency, respectively),
$\dot{M}_{\rm Edd}$ is the mass accretion rate in Eddington units and
$q$ is the mass ratio between the two coalescing black holes.

By construction, the recoil velocities that can be studied in our
setup are those contained in the equatorial (\ie $(r, \phi)$) plane
and because it is much more advantageous to study the dynamics of the
disc in a reference frame comoving with the black hole, we impose a
net velocity field in addition to the equilibrium orbital one. In
practice, at time $t=0$ we perform a Lorentz boost of the fluid
velocity along the radial direction with $\phi=0$, thus mimicking a
recoil velocity of the black hole in radial direction but with
$\phi=\pi$. We treat the imparted recoil velocity $V_{\rm k}$
essentially as a free parameter ranging from $V_{\rm
  k}=100\,\rm{km}/\rm{s}$ to $V_{\rm k}=3000\,\rm{km}/\rm{s}$, where
the latter values are not realistic and serve here only to appreciate
the disc dynamics under extreme conditions. We recall, in fact, that
the recoil velocities in the orbital plane are expected to be
$\lesssim 450\,\rm{km}/\rm{s}$~\citep{Koppitz-etal-2007aa, Herrmann:2007ac,
  Pollney:2007ss:shortal}.

In addition to the recoil, in some initial models we also consider the
effects of the mass lost to gravitational waves and which we account
by first computing the initial model in the gravitational potential of
the full black hole mass, and then evolving it in the gravitational
potential of the reduced mass. As a reference value we consider a
decrease in the mass of $\sim 3\%$, thus corresponding to that
obtained from the typical merger of equal-mass spinning black holes
with spins anti-aligned with the orbital angular momentum
~\citep[see][Fig.~$11$]{Reisswig:2009vc}. 
Higher values of the mass loss
do not introduce qualitative changes in the overall dynamics.

A final comment is devoted to the EOS of the initial model, that we
chose to be that of a polytrope $p=\kappa \rho^\gamma$, with
$\gamma=4/3$ or $\gamma=5/3$. We recall that a peculiar property of
these equilibrium models is that the ratio $p/\rho$, and therefore the
temperature $T$ and the sound speed $c_s$, do not depend on the
polytropic constant $\kappa$\footnote{The argument consists in proving
  that the function $\Psi \equiv \kappa(n+1)\Theta$, where
  $\gamma=1+1/n$ and $\rho=\Theta^n$, does not depend on
  $\kappa$. From this it follows that $p/\rho=\kappa\Theta$ does not
  depend on $\kappa$ either.}, which, on the other hand, determines
the mass of the disc~\citep[see][Appendix B]{Rezzolla_qpo_03b}.  
As a result, the size of the torus is
fixed, also the temperature is uniquely determined and cannot be
rescaled further. The last column in Table~\ref{tab1} reports such a
temperature at the centre of the torus, $r_{\rm c}$, as estimated from
the ideal-gas EOS via the expression
\be
\label{t_estimate}
T=\frac{m_p}{k_B}\frac{p}{\rho},
\ee
where, we recall, $k_B$ is the Boltzmann constant and $m_p$ the
rest-mass of the proton. In geometric units and with $m_p/k_B=1$ the
transformation of the temperature from the dimensionless values to
Kelvin degrees is given by
\be
T = 1.088\times 10^{13}\left(\frac{p}{\rho}\right) \, \, \, K \ .
\ee
%

\begin{table*}
\begin{center}
\caption{Main properties of the initial models. From left to right the
  columns report the name of the model, the black hole spin parameter
  $a$, the mass of the black hole, the disc-to-hole mass ratio, the
  power-law index $q$ of the angular momentum distribution and the
  parameter ${\cal S}$ (only for models with non constant distribution
  of the specific angular momentum), the constant value of the
  specific angular momentum $\ell$ (only for models with constant
  distribution of the specific angular momentum) the adiabatic index
  $\gamma$, 
  the inner and the outer radius of the tours, $r_{\rm in}$ and
  $r_{\rm out}$, the radius of the maximum rest-mass density $r_{\rm c}$,
  the orbital period at the radius of maximum rest-mass
  density $\tau_{\rm c}$, the maximum temperature $T_{\rm c}$, the
  orbital velocity $|v^{\phi}|=(v_{\phi}v^{\phi})^{1/2}$ at $r_{\rm
    in}$.
}
\label{tab1}
\begin{tabular}{l|ccccc|cccc|ccccc}
\hline
\hline
Model & $J/M^2$     & $M$ & $M_{\rm d}/M$ & $ q $ & ${\cal S}$ & $\ell$ & $\gamma$ & $r_{\rm in}$ & $r_{\rm   out}$ & $r_{\rm c}$ & $\tau_{\rm c}$ & $T_{\rm c}$ & $|v^{\phi}|_{\rm in}$ \\
      &     & $(M_{\odot})$ &             &       &
&        &          &            & $(M)$       & $(M)$
& $(M)$        & $(K)$ & $({\rm km/s})$\\

\hline

\texttt{S.00} & $0.00$ & $1.0\times 10^{6}$ & $2.0\times 10^{-3}$ & $-$ & $-$ & $8.0$ & $4/3$ & $40.0$ & $118.2$ & $59.8$ & $3.97$ (h)& $5.4\times 10^{9}$&$57000$\\
\texttt{S.25} & $0.25$ & $1.0\times 10^{6}$ & $2.1\times 10^{-3}$ & $-$ & $-$ & $8.0$ & $4/3$ & $40.0$ & $120.0$ & $60.0$ & $4.00$ (h)& $5.6\times 10^{9}$&$57000$\\
\texttt{S.50} & $0.50$ & $1.0\times 10^{6}$ & $2.2\times 10^{-3}$ & $-$ & $-$ & $8.0$ & $4/3$ & $40.0$ & $121.7$ & $60.2$ & $4.02$ (h)& $5.7\times 10^{9}$&$57000$\\
\texttt{S.75} & $0.75$ & $1.0\times 10^{6}$ & $2.1\times 10^{-3}$ & $-$ & $-$ & $8.0$ & $4/3$ & $40.0$ & $123.5$ & $60.4$ & $4.04$ (h)& $5.8\times 10^{9}$&$57000$\\
\texttt{S.99} & $0.99$ & $1.0\times 10^{6}$ & $2.1\times 10^{-3}$ & $-$ & $-$ & $8.0$ & $4/3$ & $40.0$ & $125.2$ & $60.6$ & $4.06$ (h)& $5.9\times 10^{9}$&$57000$\\

\hline

\texttt{L.00} & $0.00$ & $1.0\times 10^{6}$ & $1.0\times10^{-4}$ & $0.4950$ & $1.037$ & $-$ & $4/3$ & $984.6$ & $403.6$ & $4713.7$ & $11.06$ (d)& $8.7\times 10^{6}$&$14970$\\
\texttt{L.00.MA} & $0.00$ & $1.0\times 10^{6}$ & $1.0\times10^{-4}$ & $0.4950$ & $1.037$ & $-$ & $5/3$ & $984.6$ & $403.6$ & $4713.7$ & $11.06$ (d)& $1.4\times 10^{7}$&$14970$\\
\texttt{L.90} & $0.90$ & $1.0\times 10^{6}$ & $1.0\times 10^{-4}$ & $0.4955$ & $1.033$ & $-$ & $4/3$ & $988.4$ & $400.9$ & $4760.0$ & $11.13$ (d)& $7.8\times 10^{6}$&$15030$\\

\hline
\hline

\end{tabular}
\begin{flushleft}

\end{flushleft}
\end{center}
\end{table*}

\section{Methodology of the analysis}
\label{Monitored_quantities}

In what follows we discuss in detail the physical quantities computed
during the evolution either as representatives of the global evolution
or of the local one.

\subsection{Global quantities}
\label{Global_quantities}

In addition to the local Eulerian fluid variables, during the
evolution we also monitor a few global quantities that are very
helpful for interpreting the main properties of the dynamics.  These
are: the rest-mass, the internal energy and the accretion rate at the
innermost radial point of the grid, each of which is computed as
\bea
\label{mass}
&&M_{{\rm disc}} \equiv  2 H\int \sqrt{\gamma} D dr d\phi ,      \\
\label{int_enrgy}
&&E_{\rm int} \equiv 2 H \int\epsilon\sqrt{\gamma} D dr d\phi , \\
\label{dotm}
&&\dot{M}(r_{\rm min}) \equiv - 2 H \int \alpha \sqrt{\gamma}  D v^r d\phi \ .
\eea
Note that when computing the volume integral we consider the discs to
have half thickness $H$ which is assumed to be constant in radius, \ie
with $H\sim c_s/\Omega$ as in the standard thin disc approximation,
with $c_s$ the sound speed and $\Omega$ the orbital velocity.

In addition to~(\ref{mass})--(\ref{dotm}) we also compute a few more
diagnostic quantities that, on the contrary, rely on simplified
assumptions reflecting the fact that the implementation does not
account for processes such as radiation transfer and viscous
dissipation. 
In particular, we compute the bremsstrahlung emissivity
of the electron-proton collision as~\citep{Rybicki_Lightman1986}
\bea
\epsilon_{{\rm BR}}&\simeq& 2.0\times 10^{-27}T^{1/2}Z_i^2n_e n_i \ \ {\rm erg}
\ \ {\rm cm}^{-3} \ \ {\rm s}^{-1} \\
&\simeq& 7.14\times 10^{20}T^{1/2}\rho_{{\rm cgs}}^2 \ \ {\rm erg}
\ \ {\rm cm}^{-3} \ \ {\rm s}^{-1} ,
\eea
where $n_e$ and $n_i\simeq n_e$ are the number densities of electrons
and ions (protons), respectively, while $T$ is the equilibrium
temperature of both electrons and protons. The bremsstrahlung
luminosity is then obtained after performing the volume integral
\be
\label{brem_geo}
L_{{\rm BR}}\simeq 3 \times 10^{78} \int
\left( T^{1/2}\rho^2 \Gamma \sqrt{\gamma}d^3x\right)  
\left(\frac{M_\odot}{M}\right)
\ \ {\rm erg}/{\rm s} \ ,
\ee
where the large multiplicative factor comes from the fact that both
$T$ and $\rho$ in~(\ref{brem_geo}) are expressed in geometrized units.

\subsection{A relativistic ``shock detector''}
\label{Shock_detection}

An obvious expectation, which has been confirmed by all of the
numerical simulations to date, is that the as the recoiling black hole
will move in the plane of the accretion disc it will introduce spiral
shocks which will move outwards on a timescale which is comparable
with the orbital one. Because determining the accurate position of the
shocks is important to correlate the latter to the EM emission, a
number of suggestions have been made in the literature, which have a
varying degree of precision. In particular,~\citet{Lippai:2008,
  Oneill2009, Megevand2009}, all just looked at density and/or
pressure gradients to infer the propagation of a spiral caustic and,
therefore, of a possible shock (we note that in the collisionless
particles treatment of~\citet{Lippai:2008}, the existence of a shock
is purely indicative as no shocks can be produced in this
approximation). On the other hand, \citet{Rossi2010} used the
introduction of an artificial viscosity, which is itself related to
local density increases, to identify the location of shocks.
Finally,~\citet{Corrales2009} used a shock detector present in the
\texttt{FLASH} code, which marks a given region as a shocked one if
$\vec{\nabla} \cdot \vec{v} < 0$ and if the pressure difference
between the monitored zone and at least one of its neighbors exceeds
the difference expected from the Rankine-Hugoniot jump condition for a
shock of a pre-specified minimum Mach number. While more robust than
those considered by the other authors, also this prescription is a
delicate one as we will discuss in Sec.~\ref{Shock_propagation}.

All of the methods mentioned above contain rather empirical
  criteria and can fail to detect shocks unless they are
very strong. To improve the determination of the location of the
shock, even when the latter are arbitrarily weak, we have devised a
relativistic ``shock detector'' which exploits an idea discussed in
all its details in~\citet{Rezzolla02} and~\citet{Rezzolla03}, and
which consists essentially in the possibility of predicting the
outcome of the wave pattern in a Riemann problem. (We note that a
similar detector can be prescribed also for non-relativistic flows;
the interested reader can find a detailed discussion in \textsection
100 of~\citet{Landau-Lifshitz6}.).

To illustrate the logic of our shock detector let us suppose that
along a given direction, say the $x-$direction, two adjacent fluid
elements $1$ and $2$ manifest a jump in the hydrodynamical quantities,
such as pressure, density and velocity, thus reproducing the typical
conditions of a local Riemann problem.  In the absence of magnetic
fields, the time evolution of a Riemann problem consists in the
propagation along opposite directions of two nonlinear waves, either
rarefactions or shocks, separated by a third wave, the contact
discontinuity.  As a result, a shock front will be produced if the
wave pattern generated by the Riemann problem contains at least one
shock wave, while the other wave can be a rarefaction wave.  As shown
by~\citet{Rezzolla02}, there is a simple criterion for predicting the
occurrence of a wave pattern containing a shock wave and this amounts
to the requirement that the relative velocity between the two states
$1$ and $2$ (\ie between two adjacent fluid cells) is larger than a
threshold value
\be
\label{condition}
v_{12}>({\tilde v}_{12})_{_{SR}},
\ee
where $({\tilde v}_{12})_{_{SR}}$ is a function of the thermodynamic
states of $1$ and $2$, while $v_{12}\equiv(v_1-v_2)/(1-v_1 v_2)$ is
the special relativistic expression for the relative velocity. When
there are nonzero velocities in the direction tangential to the
discontinuity front, the analytic form of $({\tilde v}_{12})_{_{SR}}$
is given by~\citep{Rezzolla03}
\begin{equation}
\label{capo2_analytic}
(\widetilde{v}_{12})_{_{SR}}\equiv
	\tanh\left(\int_{p_1}^{p_2} \frac{\sqrt{h^2 + 
	{\cal A}^2_1(1-c_s^2)}}
        {(h^2 + {\cal A}^2_1)\rho~c_s} dp \right)\ ,
\end{equation}
where ${\cal A}_1 \equiv h_1 W_1 v^y_1$ while $c_s$ is the sound
speed. If, on the contrary, the relative velocity $v_{12}$ is smaller
than $({\tilde v}_{12})_{_{SR}}$, then no shock wave can be produced
and the wave pattern of the corresponding Riemann problem consists of
two rarefaction waves propagating in opposite directions.

With this idea in mind we have constructed a sensitive shock detector
to locate the regions of the disc which are producing a spiral
shock. In practice we first select the direction along which we want
to monitor the propagation of shock waves. Secondly, since
(\ref{condition}) and~(\ref{capo2_analytic}) have been derived in a
flat spacetime, we project the velocity field $v^j$ in a local tetrad
in Boyer-Lindquist coordinates so as to obtain the new components
$v^{\hat j}$
\bea
\label{tetrad1}
&&v^{\hat r}=\sqrt{g_{rr}}v^r , \\
\label{tetrad2}
&&v^{\hat \phi}=\sqrt{g_{\phi\phi}}v^\phi \ .
\eea
Thirdly, we calculate $v^{\hat x}$ and $v^{\hat y}$ from $v^{\hat r}$
and $v^{\hat \phi}$ through a simple rotation. Finally, we compute the
integral (\ref{capo2_analytic}) in terms of the hatted Cartesian
components and compare the result with $v_{12}$.

Note that the integral~(\ref{capo2_analytic}) effectively provides the
minimum value for the occurrence of a wave pattern containing a single
shock wave. In the limit of $(\widetilde{v}_{12})_{_{SR}}\rightarrow
v_{12}$, in fact, the pressure jump across the shock wave becomes
vanishingly small and a single rarefaction wave joining $p_1$ and
$p_2$ propagates in the direction opposite to that of the vanishing
shock wave. Therefore, when computing~(\ref{capo2_analytic}) we are
actually integrating inside the rarefaction wave, that is notoriously
a self-similar solution and hence isentropic. This means that in
evaluating~(\ref{capo2_analytic}) we can use the isentropic expression
for the sound speed
\be
\label{sound_speed}
c_s=\sqrt{\frac{\gamma(\gamma-1)p}
	{(\gamma-1)\rho+\gamma p}}\ ,
\ee
where the density $\rho$ is given in terms of $p$ from $p=p_1
(\rho/\rho_1)^\gamma$.

\begin{figure*}
\centering
{\includegraphics[angle=0,width=8.3cm,height=7.5cm]{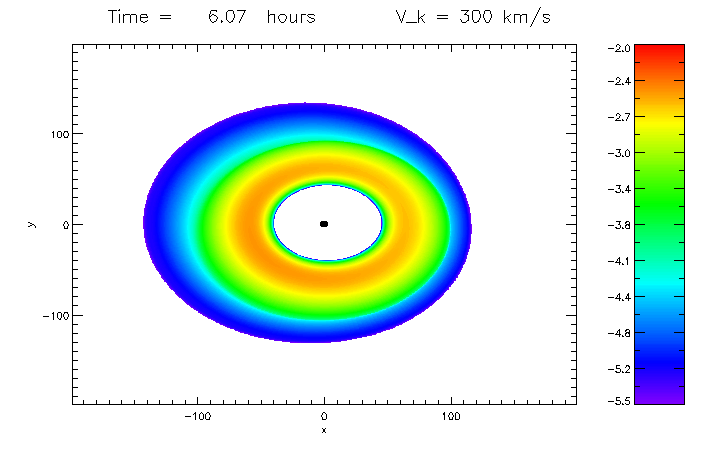}}
{\includegraphics[angle=0,width=8.3cm,height=7.5cm]{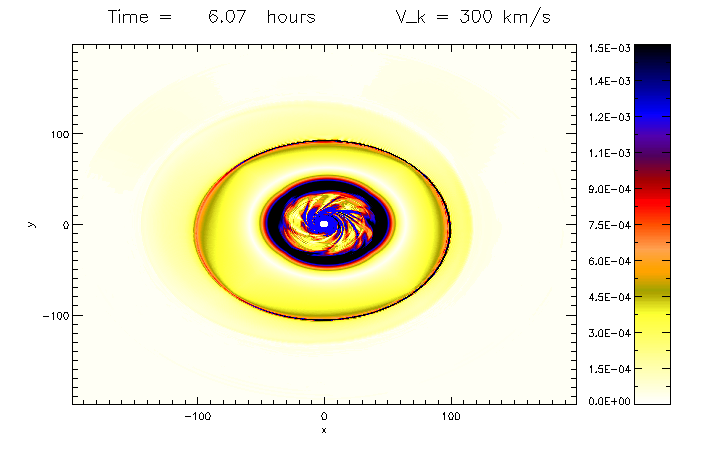}}
{\includegraphics[angle=0,width=8.3cm,height=7.5cm]{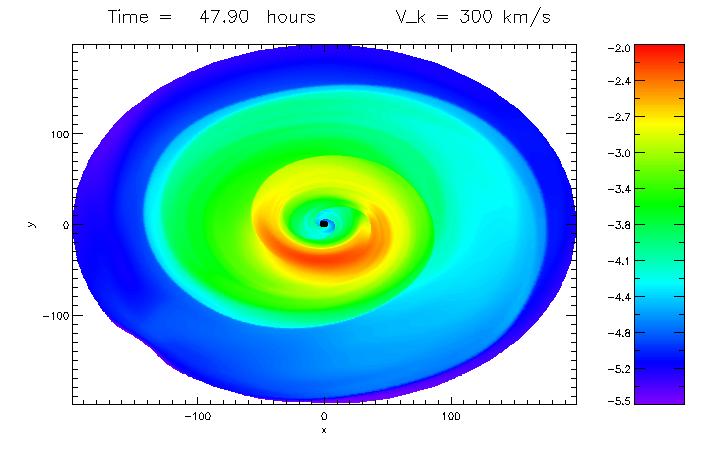}}
{\includegraphics[angle=0,width=8.3cm,height=7.5cm]{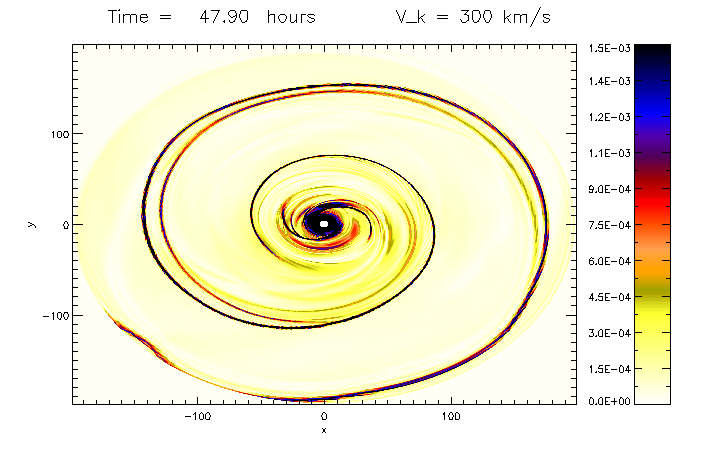}}
{\includegraphics[angle=0,width=8.3cm,height=7.5cm]{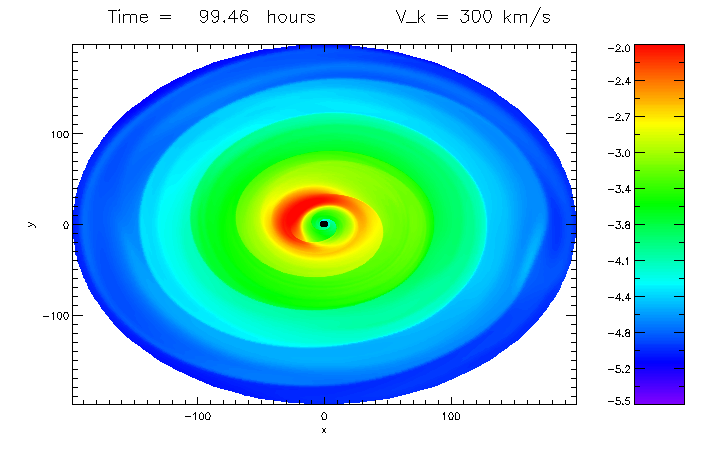}}
{\includegraphics[angle=0,width=8.3cm,height=7.5cm]{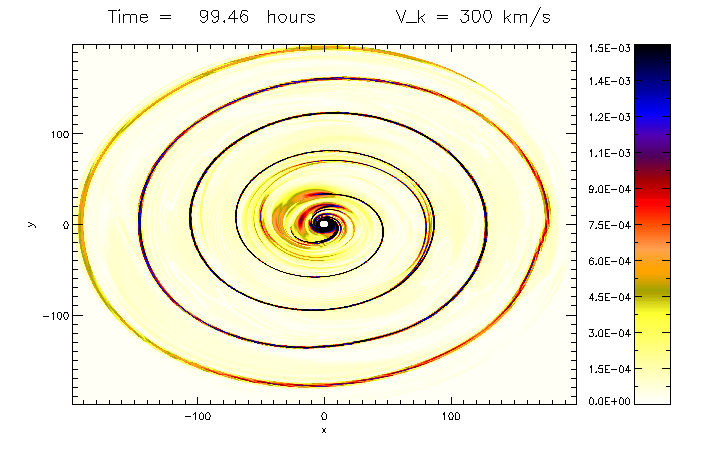}}
\caption{Rest-mass density distributions (left columns) and shock
  structure (right columns) and at three different times (\ie $t=6.07,
  47.90$ and $99.46\,{\rm h}$) for model \texttt{S.00} and a recoiling
  velocity $V_{\rm k}=300\, {\rm km/s}$. Note that the last panel
  refers to almost $25$ orbital revolutions. The rest-mass density is
  plotted on logarithmic scale and in ${\rm cgs}$ units, while the
  shock structure is obtained by plotting the quantity $S_d$ (see
  beginning of Sec.~\ref{Shock_propagation} for a definition); shock
  waves can form in regions where $S_d>0$. Note that is very hard to
  locate a shock by simply looking at the density distribution.}
\label{fig1}
\end{figure*}

\begin{figure*}
\centering
{\includegraphics[angle=0,width=8.3cm,height=7.5cm]{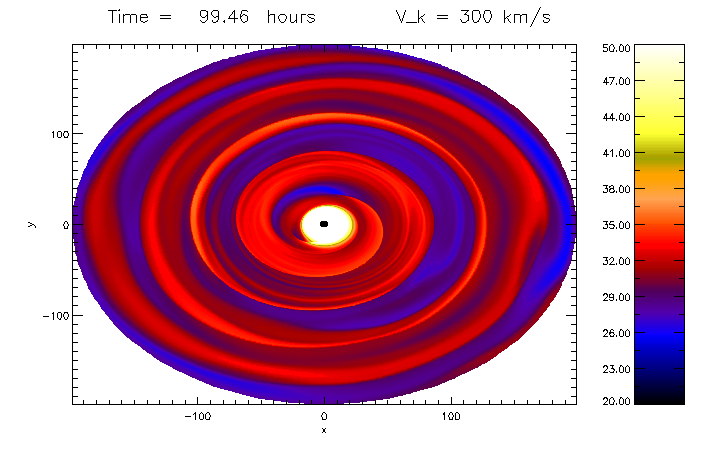}}
{\includegraphics[angle=0,width=8.3cm,height=7.5cm]{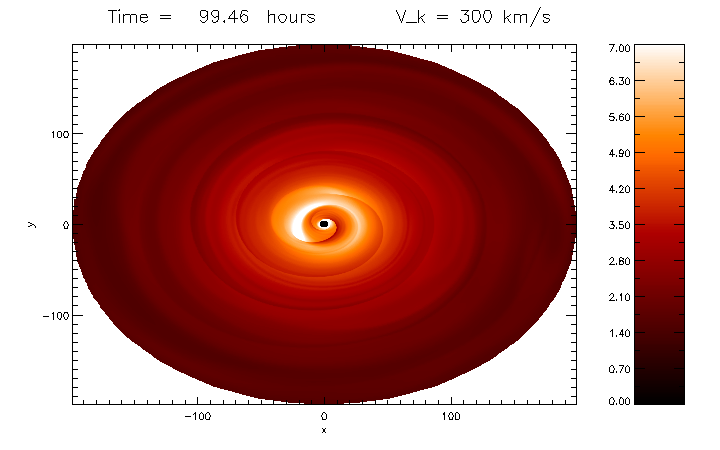}}
\caption{Sound speed normalized to the kick velocity $c_s/V_{\rm k}$
  (left panel) and relativistic Mach number ${\cal M}$ (right panel)
  at $t=99.46\,{\rm h}$ for model \texttt{S.00} when subject to a
  recoil of $V_{\rm k}=300\, {\rm km/s}$. Note that $V_{\rm k} \geq
  c_s$ is not a good criterion for the localization of
  the shock (\cf
  left panel) and that no obvious correlation is present between the
  supersonicity of the flow and the appearance of the
  shock (\cf right
  panel). 
}
\label{fig2}
\end{figure*}

The procedure described above is completely general and can be
proposed as an efficient shock detector for numerical relativistic
hydrodynamics. However, two subtleties should also to be taken into
account. The first subtlety is that, for more complicated spacetimes
or coordinates systems, the flat-spacetime
projection~(\ref{tetrad1})--(\ref{tetrad2}) should be replaced by the
more general form
\begin{equation}
v^{\hat{i}}=M_{~j}^{\hat{i}} v^j,
\end{equation}
with $M_{~j}^{\hat{i}}$ given by~\citep{Pons1998}
\begin{equation}
\displaystyle{
M_{~j}^{\hat{i}} = \left( 
\begin{array}{ccc}
\sqrt{\gamma_{11}} & \frac{-\gamma^{12} \gamma_{22}+\gamma^{13} \gamma_{23} }{\gamma^{11} \sqrt{\gamma_{22}}} & \frac{-\gamma^{13} \sqrt{\gamma_{22} \gamma_{33}-(\gamma_{23})^2} }{\gamma^{11} \sqrt{\gamma_{22}}} \cr
&   &  \cr
0   &  \sqrt{\gamma_{22}}    &  0   \cr
&   &  \cr
0   &  \frac{\gamma_{23}}{\sqrt{\gamma_{22}}} & \frac{\sqrt{\gamma_{22} \gamma_{33}-(\gamma_{23})^2} }{\sqrt{\gamma_{22}}}  \cr
\end{array} 
\right) \ .
}
\end{equation}
%
The second subtlety concerns the fact that, because the shock detector
validates the inequality~(\ref{condition}), it can be arbitrarily
sensitive. Although this certainly represents an advantage, one often
wishes to disregard the whole class of weak shocks, for which the
contribution to the entropy jump is of higher order and $\Delta
s\propto (\Delta p)^3$~\citep{Thorne73}. In these cases the weakest
shocks can be filtered out by making the condition~(\ref{condition})
somewhat more restrictive and require therefore that a shock is
detected if
\be
\label{condition2}
v_{12}>{\tilde v}_{12}=({\tilde v}_{12})_{_{SR}}+\chi\left[({\tilde
    v}_{12})_{_{2S}} - ({\tilde v}_{12})_{_{SR}}\right]
\ee
where 
\be
(\widetilde{v}^x_{12})_{_{2S}}
  =\frac{(p_1-p_2)(1-v^x_2 \bar{V}_s)}
	{(\bar{V}_s-v^x_2)\{h_2 \rho_2
	(\Gamma_2)^2 [ 1-(v^x_2)^2 ] + p_1 - p_2\}}  \ ,
\ee
with $\bar{V}_s$ being the velocity of the shock wave propagating
towards state $2$~\citep[see][for the explicit
expression]{Rezzolla03}. 
Because $({\tilde v}_{12})_{_{2S}} \geq ({\tilde
  v}_{12})_{_{SR}}$, any value of $\chi$ between $0$ and $1$ will
effectively raise the threshold for the detection of the shocks,
filtering out the weakest ones; the shocks encountered in the
simulations reported here were all rather weak and we have therefore
always used $\chi=0$. The whole procedure is repeated for as many
directions as the dimensions of the problem.  Finally, a prescription
of the relativistic shock detector as adapted for Newtonian fluids is
presented in Appendix~\ref{appendixA}.

\section{Results}
\label{Results}

\subsection{Small-size models}
\label{Small_size_models}

Although the small-size models are not astrophysically very realistic
as they presume the existence of small tori in equilibrium near the
recoiling black hole, they serve to set a comparison with the other
general-relativistic calculations of~\citet{Megevand2009}, where
similar tori were considered. In addition, by being so close to the
black hole, they are helpful in capturing those features of the
dynamics that are most influenced by the regions of strong gravity.
However, because of their limited extensions and high
densities/temperatures (as an example, the model \texttt{S.00} has
$\rho_{{\rm c}}=3.38\times 10^{-3}{\rm g/cm^3}$ and $T_{{\rm
    c}}=7.9\times 10^8 K$) they will not be used to draw any
conclusion on the emitted luminosity, which will be instead discussed
in more detail when analyzing the large-size models in
Sec.~\ref{Large_size_models}.

\subsubsection{Shock Dynamics}
\label{Shock_propagation}

The different panels in the left column of Fig.~\ref{fig1} show the
rest-mass density at three different times (\ie $t=6.07, 47.90$ and
$99.46\,{\rm h}$) for model \texttt{S.00} and a recoiling velocity
$V_{\rm k}=300\, {\rm km/s}$. Although the imparted velocity is rather
small (but close to the maximum possible in the orbital plane), the
disc undergoes large variations in size and density, with a shock
front that expands from the inner parts of the disc in an initially
axisymmetric manner. This is essentially due to the reduction in the
black-hole mass and which moves all of the equilibrium orbits to
larger radii. As the influence area of the black hole becomes larger
and the orbital velocities become comparable with that of the recoil,
the disc develops shocks with the characteristic spiral structure
discussed also in previous works~\citep{Lippai:2008, Corrales2009,
  Rossi2010} and that transports angular momentum outwards. This is
shown by the panels in the right column of Fig.~\ref{fig1}, which
report the location of the shocks as obtained with the procedure
illustrated in Sec.~\ref{Shock_detection}. More specifically, they
show the quantity $S_d \equiv {\rm {max}}\{0,v^x_{12} -
\widetilde{v}^x_{12}, v^y_{12} - \widetilde{v}^y_{12}\}$, whereby any
positive value of $S_d$ marks a shocked region. Note that the region
very close to the black-hole horizon, namely at radii smaller than
$\sim 10 M$, is always a highly shocked one. Furthermore, as the
evolution proceeds and the disc expands first in response to the mass
loss and subsequently to the shocks, very little (if any) of the
computational domain is filled by atmosphere, thus removing de-facto
any role it can play in the dynamics of the disc.

Figure~\ref{fig2} provides additional information about the dynamics
of the disc by showing the local sound speed normalized to the kick
velocity, and the relativistic Mach number, the latter computed as
${\cal M}\equiv \Gamma v/\Gamma_s c_s$, where $\Gamma_s\equiv
1/\sqrt{1-c_s^2}$ is the Lorentz factor of the sound
speed~\citep{Konigl1980}. Our results shows that the criterion
suggested by ~\citet{Corrales2009} for the occurrence of shocks,
namely $V_{\rm k}\geq c_s$, can be rather misleading in the
relativistic context. Indeed, as shown by the snapshots at time
$t=99.46\,{\rm h}$ in Fig.~\ref{fig1}, and which refers to almost $25$
orbital revolutions, a clear spiral shock forms even if the sound
speed is more than $30$ times larger than the kick velocity. Indeed,
the left panel of Fig.~\ref{fig2} seems to suggest that if any,
$c_s/V_{\rm k} \gg 1$ is possibly a reasonable necessary condition for
the approximate location of the shock. In addition, the correlation
between the occurrence of a shock and the local sound speed is very
weak and this is apparent in the right panel of Fig.~\ref{fig2}, where
it is clear that the flow is highly supersonic in the inner regions of
the disc and mildly subsonic in the outer regions. Yet, the
spiral-shock structure extends continuously across the whole disc. We
also note that although the precise morphology of the spiral shocks
will depend on the spin of the black hole, this dependence is only
very weak and all the considerations made above for model
\texttt{S.00} hold true qualitatively also for spinning black holes.

It is finally worth remarking that the shocks formed here are very
mild and not relativistic. Even for $V_{\rm k}=3000\, {\rm km/s}$, the
shock velocity maintains a typical value $V_s\sim 0.15$ and the
velocity jump at the shocks is also rather limited,
producing a $v/\Delta v\sim 30$. 
This means, for instance, that such shocks
are unable of accelerating electrons through the classical mechanism
of \citet{Bell1978}.

\begin{figure}
{\includegraphics[angle=0,width=9.0cm]{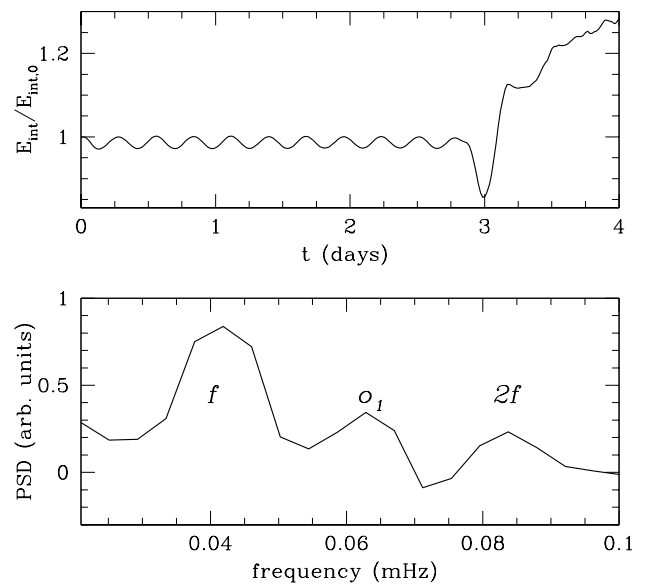}}
\vspace{-2.0cm}
\caption{Time evolution of the internal energy when normalized to the
  its initial value (top panel) and the corresponding power spectrum
  (bottom panel) in a model with $V_{\rm k}=0$ and with a mass loss of
  $1\%$ the initial mass of the black hole. }
\label{fig4}
\end{figure}

\subsubsection{Mass loss and Quasi-Periodic Dynamics}

The dynamics of the disc can change considerably if the black hole is
assumed to be recoiling with negligible velocity in the orbital plane
and only mass loss is taken into account.  By considering mass losses
in the range $1\%-10\%$, ~\citet{Oneill2009} showed that shocks can
form even in the absence of a recoil velocity, provided that the mass
loss is larger than the half thickness of the disc. The perturbation
induced by the mass loss is spherically symmetric and it causes the
disc to expand as each fluid element will want to move to the larger
radii corresponding to the equilibrium orbit for the given initial
angular momentum. Together with this expansion, however, restoring
forces will also induce the disc to contract in the
effective-potential well of the black hole with the characteristic
frequency of the lowest order $p$-mode, and which is not too different
from the epicyclic frequency at the disc
centre~\citep{Rezzolla_qpo_03b}. We recall that the restoring force
responsible for the appearance of such $p$ modes is a combination of
pressure gradients, centrifugal and gravitational forces, with the
last two playing the dominant roles for the discs considered
here~\citep{Kato2001,Rezzolla_qpo_03b}.

\begin{figure}
{\includegraphics[angle=0,width=9.0cm]{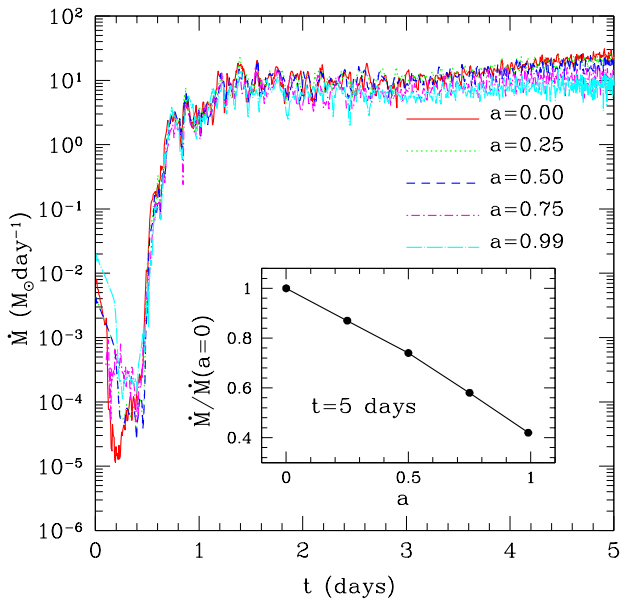}}
\vspace{-2.0cm}
\caption{Mass accretion rate measured at $r=r_{\rm{min}}$ for
  different values of the black hole spin-parameter in the small-size
  models, \ie \texttt{S.00}--\texttt{S.99}. Shown in the inset as a
  function of the black-hole spin is the stationary accretion rate
  reached after $5\,{\rm d}$.}
\label{fig3}
\end{figure}

The oscillating behavior induced by the sudden change of the potential
well and the subsequent development of the instability is shown in
Fig.~\ref{fig4}, where the top panel reports the time evolution of a
typical global quantity, \ie the internal energy, when normalized to
its initial value.  Interestingly, the remarkable periodicity that
characterizes the dynamics is the same as found by~\citet{Zanotti03}
when studying global modes of oscillation of thick discs around black
holes. The corresponding power spectrum is shown in the bottom panel
of Fig.~\ref{fig4}, obtained through a FFT of the time series for
$t\lesssim 3\,{\rm d}$ reveals the presence of a fundamental mode of
oscillations at $f\sim 4.17\times 10^{-5}\, {\rm Hz}$ and of two
overtones. The first overtone is at $o_1\sim 6.28\times 10^{-5}\, {\rm
  Hz}$, while the second one, very close to twice the fundamental
frequency, $o_2\sim 8.37\times 10^{-5}\, {\rm Hz}\sim 2 f$, is
produced by nonlinear coupling of the fundamental mode
with itself~\citep[see][]{Zanotti05}.  
Collectively, these modes
of oscillations provide a series of modes in the same ratio of the
integer numbers $2:3:4$ observed in the QPOs of low-mass X-ray
binaries containing a black hole~\citep[see][for a
  recent review]{Remillard2006} 
and for which a simple model based on basic $p$-mode
oscillations of a small accretion torus orbiting close to the black
hole was recently proposed~\citep{Rezzolla_qpo_03a}, and which remains
one of the most convincing explanations of the observed
phenomenology~\citep{Schnittman06}.

\begin{figure}
{\includegraphics[angle=0,width=9.0cm]{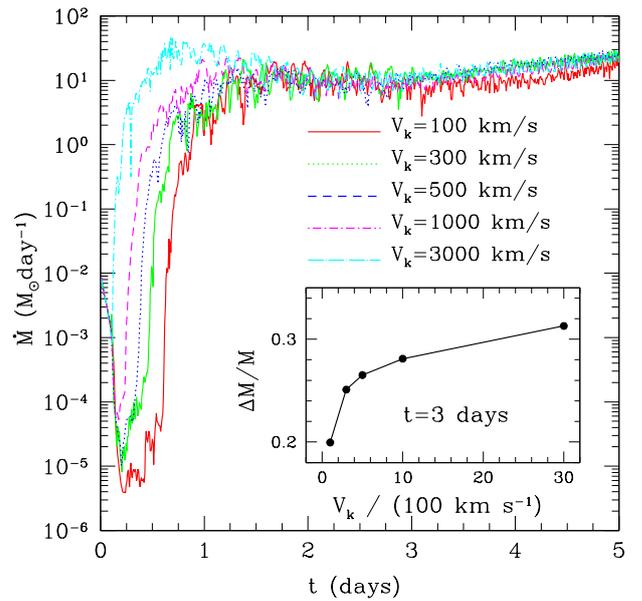}}
\vspace{-2.0cm}
\caption{Mass accretion rate measured at $r=r_{\rm{min}}$ for the
  small-size model \texttt{S.00} and different values of the recoil
  velocity. Shown in the inset as a function of the recoil velocity is
  the relative baryon mass accreted after $3\,{\rm d}$.}
\label{fig5}
\end{figure}

It is difficult not to note that the harmonic behaviour shown in the
top panel of Fig.~\ref{fig4} is lost at $t\simeq 3\,{\rm d}$ and that
the internal energy increases monotonically after that. This is due to
the onset of a non-axisymmetric instability which produces spiral arms
that rapidly spread to cover the whole disc. An instability of this
type was not pointed out by~\citet{Megevand2009} although they have
used similar models and we believe that this is probably because their
simulations were interrupted after $\sim 11\,{\rm h}$ (or $\sim 6$
orbital periods as measured at the point of maximum rest mass
density), which is too early for the development of the
instability. On the other hand, such type of instabilities in
non-Keplerian discs have been discussed by a number of authors,
starting from the pioneering work by~\citet{Papaloizou84}. A detailed
comparison between the linear perturbative analysis of these
instabilities and two-dimensional numerical simulations in a
Schwarzschild spacetime was already proposed more than twenty years
ago by~\citet{Blaes1988}, who found the development of the same spiral
structures, which transport both mass and angular momentum outwards
even in the absence of mass loss\footnote{We have verified that the
  spiral arms do indeed develop also in the absence of a mass loss or
  of a recoil and that also in this case the instability takes place
  after $\sim 4\,{\rm d}$.}.  While we cannot concentrate here on a
detailed discussion of these instabilities, it is sufficient to remark
that a spiral-shock pattern and all of the associated phenomenology,
can be generated even when the recoil velocity is zero.

\begin{figure*}
\centering
{\includegraphics[angle=0,width=8.3cm,height=7.5cm]{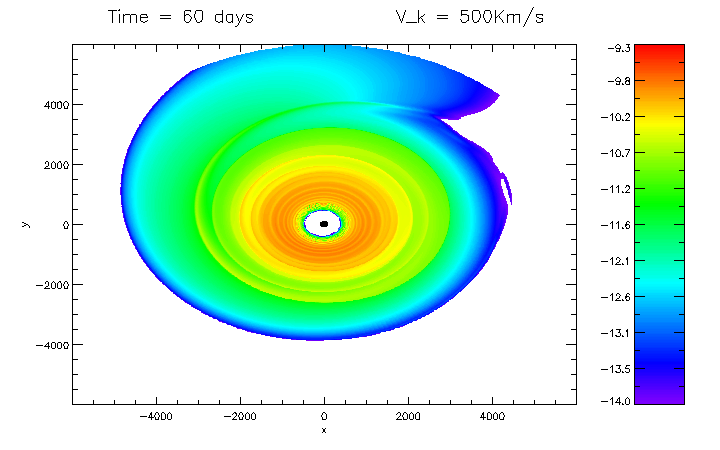}}
{\includegraphics[angle=0,width=8.3cm,height=7.5cm]{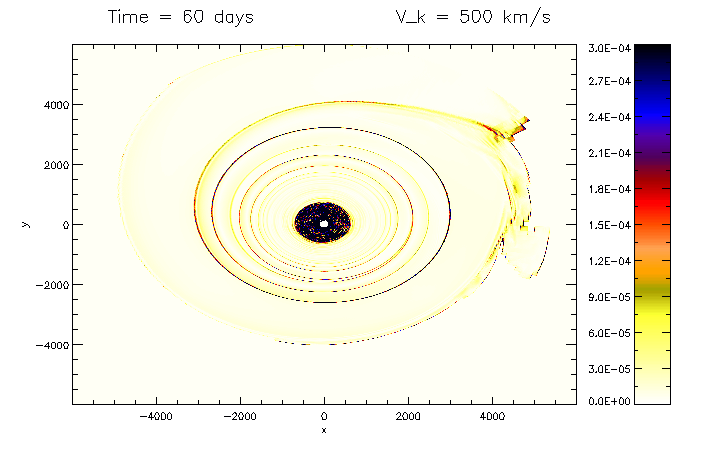}}
{\includegraphics[angle=0,width=8.3cm,height=7.5cm]{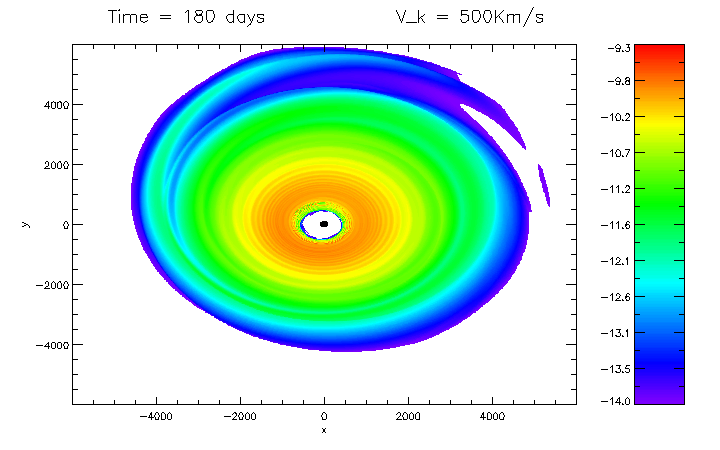}}
{\includegraphics[angle=0,width=8.3cm,height=7.5cm]{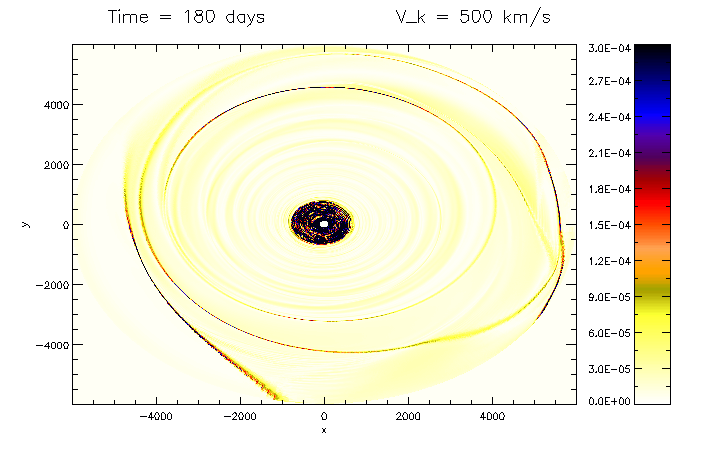}}
\caption{Left column: Rest-mass density after $60$ and $180\,{\rm d}$
  for a recoil velocity $V_{\rm k} = 500\,{\rm km/s}$ applied to model
  \texttt{L.00}.  The scale is logarithmic and expressed in ${\rm
    cgs}$ units. Right column: Shock structure as presented in
  Fig.~\ref{fig1} for the same panels in the left column. Once again
  we remark that is very hard to locate a shock by simply looking at
  the density distribution, especially when they are very weak (here
  we have used $\chi=0$ in Eq.~\ref{condition2}). Note also that the
  temperature distribution is inversionally proportional to that of
  the density (not shown here).}
\label{fig7}
\end{figure*}

Overall, the dynamics observed for model \texttt{S.00} suggests that
transient oscillating phenomena may exist in the post-merger phase of
SMBBH. In this case, the occurrence of QPOs in the accretion and thus
in the luminosity of potential EM counterparts, followed then by the
development of non-axisymmetric instabilities would be a unique and
convincing signature that a SMBBH merger with small recoil velocities
has taken place.

\subsubsection{Accretion rates}

As originally pointed out by~\citet{Kozlowski1978}, in non-Keplerian
discs no viscosity is needed in order to support accretion in the
vicinity of the cusp.  Figure~\ref{fig3} reports the baryon mass
accretion rate measured at $r=r_{\rm{min}}$ for different values of
the black hole spin-parameter in the small-size models, \ie
\texttt{S.00}--\texttt{S.99}, and provides further support to the
interpretation of the shock dynamics discussed above. In particular,
it is easy to realize that because of the sudden mass loss and hence
of the reduced gravitational attraction, the disc reacts to the excess
of angular momentum by expanding. This effect only lasts for a couple
of orbital periods after the merger and leads to the large decrease in
mass accretion rate shown in Fig.~\ref{fig3} for $t \lesssim 0.6\,{\rm
  d}$.

Subsequently, as the effect of the kick velocity extends to regions of
the flow with smaller orbital velocities and becomes dominant,
non-axisymmetric density structures form and the perturbed disc starts
filling the low density central cavity while increasing the accretion
rate. This is reflected by the the large increase in the mass
accretion rate shown in Fig.~\ref{fig3} for $t \gtrsim 0.6\,{\rm d}$,
which is essentially independent of the black-hole spin.

\begin{figure*}
\centering
{\includegraphics[angle=0,width=8.3cm,height=7.5cm]{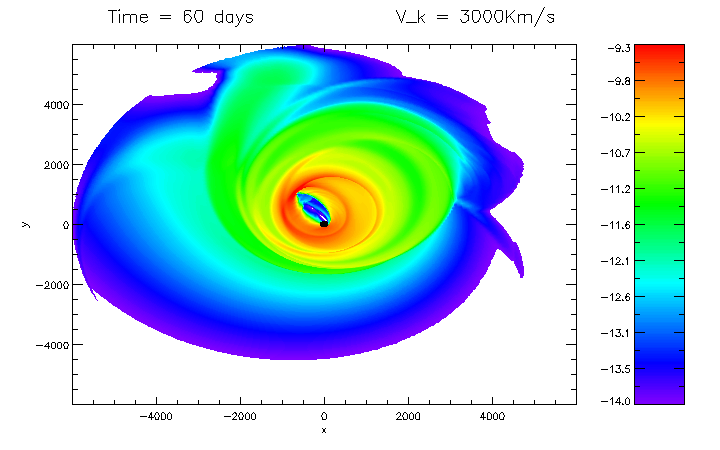}}
{\includegraphics[angle=0,width=8.3cm,height=7.5cm]{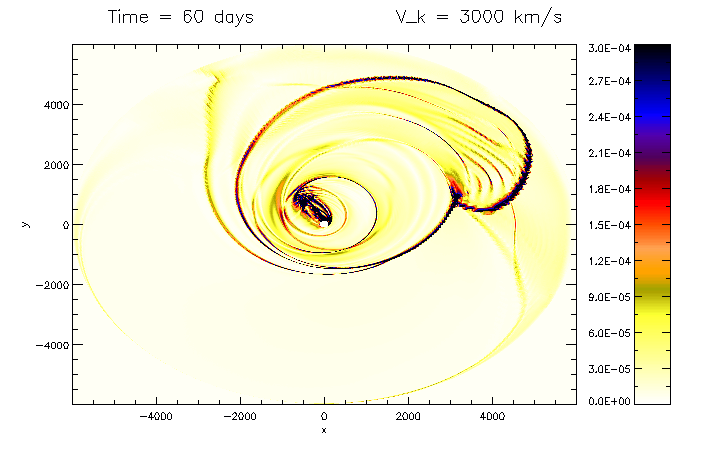}}
{\includegraphics[angle=0,width=8.3cm,height=7.5cm]{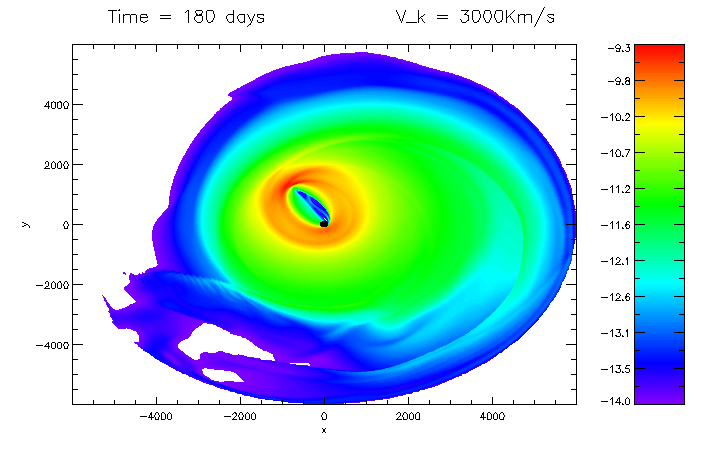}}
{\includegraphics[angle=0,width=8.3cm,height=7.5cm]{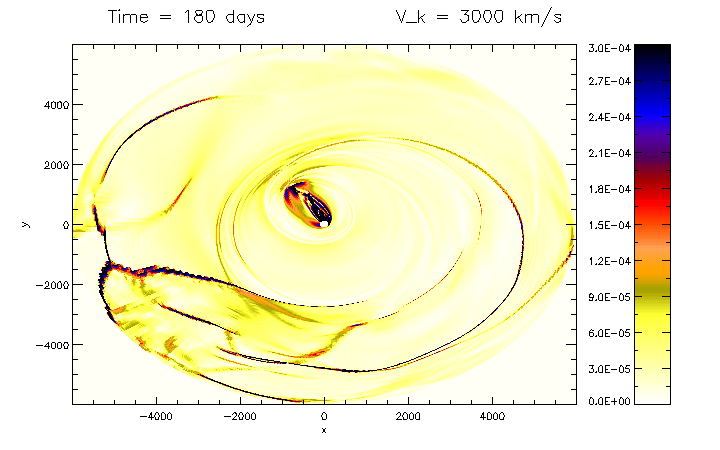}}
\caption{The same as in Fig.~\ref{fig7} but for a recoil velocity
  $V_{\rm k}= 3000\,{\rm km/s}$. Note that the spiral-shock structure
  is never present and that the inner cavity is rapidly filled by
  accreting gas. In addition, an oblique shock comprising a
  low-density region is formed in the inner parts of the
  flow. 
}
\label{fig8}
\end{figure*}

Since our treatment does not include the compensating effect of the
radiation drag exerted by the photons on the accreting matter, the
accretion rate increases undisturbed reaching values that are $\sim 6$
orders of magnitude above the Eddington limit. After about $6$ orbital
periods ($\approx 1\,{\rm d}$) $\dot M$ saturates and after $\sim
5\,{\rm d}$ all of the models have lost $\sim 32\%$ of their
mass. Fig.~\ref{fig3} also shows that the spin of the black hole has
little influence on the dynamics of the disc, although not as little
as inferred from Fig.~13 of~\citet{Megevand2009}. In particular, we
have found that after $\sim 3\,{\rm d}$ the accretion rate slightly
decreases when increasing the spin-parameter (see the inset of
Fig.~\ref{fig3}), so that $\dot{M}_{\vert a=0}\sim 2.5\,
\dot{M}_{\vert a=0.99}$ after $\sim 5\,{\rm d}$ of
evolution. Interestingly, this dependence of the accretion rate on the
spin of the black hole is rather generic and has been found also in
other simulations (not reported here) where no perturbation on the
flow is introduced.

Of course, the effect of an increasingly large kick velocity on the
disc dynamics is much more pronounced in this case, as it emerges from
Fig.~\ref{fig5}, which reports the accretion rate for different values
of $V_{\rm k}$. As already found by~\citet{Megevand2009}, larger
recoil velocities tend to anticipate the occurrence of the burst in
the accretion rate and, simultaneously, produce larger absolute values
of $\dot M$ at the burst. After the initial burst, however, and no
later than $\sim 2$ days, the accretion rates become nearly
independent of the recoil velocity. The differences in $\dot M$ are
illustrated in the inset of Fig.~\ref{fig5} which shows the relative
baryon mass accreted after $3\,{\rm d}$ and which shows a variation of
$\sim 50\%$ at most. It is also interesting to note that there is no
clear imprint of the spiral shock dynamics onto the accretion
rate. This is due to the fact that the spiral shocks essentially
redistribute angular momentum and thus modify the disc structure and
dynamics far from the black hole.

\subsection{Large-size model}
\label{Large_size_models}

We next switch to discussing the dynamics of the large-size model
\texttt{L.00} for different values of the recoil velocity. We recall
that there are at least two different reasons to consider this second
class of models. The first one is that they reflect the expectations
of a circumbinary disc: they are quasi-Keplerian, extended and at a
large distance from the binary. The second one is that by having a
lower density, \ie $\rho_{{\rm c}}=1.38\times 10^{-10}{\rm g/cm^3}$,
and hence a lower temperature, \ie $T_{{\rm c}}=8.7\times 10^6 K$,
they lead to much more reasonable values of the recoil-enhanced
luminosity. In what follows we briefly review the overall dynamics and
then concentrate on how to compute a realistic estimate for the
luminosity.

\subsubsection{Shock Dynamics}

The overall dynamics of the large-size models is qualitatively similar
to that of the smaller counterparts but with three important
differences. The first one is a much weaker dependence of the dynamics
on the recoil velocity. This is essentially due to the fact that the
inner edge of the discs is so far from the recoiling black hole that
only extremely large (and unrealistic) values of the recoil velocity
induce a significant modification of the orbital velocity. This is
shown in Fig.~\ref{fig7}, which reports in the left column the
rest-mass density after $60$ and $180\,{\rm d}$ for a recoil velocity
$V_{\rm k} = 500\,{\rm km/s}$ applied to model \texttt{L.00}. The
right column shows instead the corresponding shock structure with the
same notation used in Fig.~\ref{fig1} and highlights that a clear
spiral structure is lost already after $180\,{\rm d}$, which now
corresponds to almost $18$ orbital periods. Note that the temperature
distribution is inversionally proportional to that of the density (not
shown in Fig.~\ref{fig7}) and that the central region is filled only
with the atmosphere fluid and thus appears as white in the left
column. However, several small shocks are produced in this cavity and
these are clearly revealed by the shock detector images on the right
column which reports the central region as dark; this is just an
artifact that does not have a dynamical impact. The behaviour in
Fig.~\ref{fig7} should also be contrasted with the spiral-shock
structure of model \texttt{S.00}, which instead persisted intact after
almost $25$ orbital periods (\cf bottom right panel of
Fig.~\ref{fig1}). This behaviour and the rapid disappearance of the
spiral shock structure is further pronounced as the recoil velocity is
increased to $V_{\rm k} = 1000\,{\rm km/s}$ (not shown here).

The second difference is that for sufficiently large recoils, namely
for $V_{\rm k}\geq 2000 \,{\rm km/s}$, the initial cavity between the
black hole horizon and the inner edge of the disc can be filled
rapidly by infalling material. This is evident when looking at
Fig.~\ref{fig8}, which reports the rest-mass density and shock
structure after $60$ and $180\,{\rm d}$ for a recoil velocity $V_{\rm
  k} = 3000\,{\rm km/s}$ applied to model \texttt{L.00}. When
contrasted with Fig.~\ref{fig7}, which has the same spatial extent, it
is easy to notice that already after $\sim 60\,{\rm d}$, or $\sim 5$
orbital revolutions, the central cavity is filled with high-density
material, some of which extends right onto the black hole. Similarly
to what already seen for smaller recoils, also the right column of
Fig.~\ref{fig8} shows that in this case the spiral structure is not
present, although spatially extended shocks are formed both in the
inner regions and in the outer parts of the disc. Note that the
velocity jump at the shocks is again rather small, being at most
$\Delta v\sim 2.5\times 10^{-4}$, hence insufficient to accelerate
particles through the various acceleration mechanisms involving shock
waves.

\begin{figure}
{\includegraphics[angle=0,width=8.75cm,height=7.91cm]{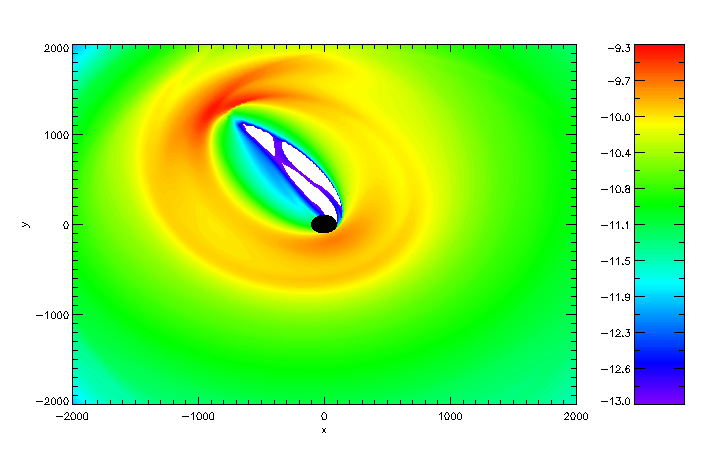}}
\caption{Magnification of the central region of the rest-mass density
  in model \texttt{L.00} after $180\,{\rm d}$ and for a recoil
  velocity $V_{\rm k}=3000 {\rm km/s}$. Note that the colormap is
  slightly different from the one in Fig.~\ref{fig8} to highlight the
  presence of a cavity.}
\label{fig12}
\end{figure}

\begin{figure}
{\includegraphics[angle=0,width=9.0cm]{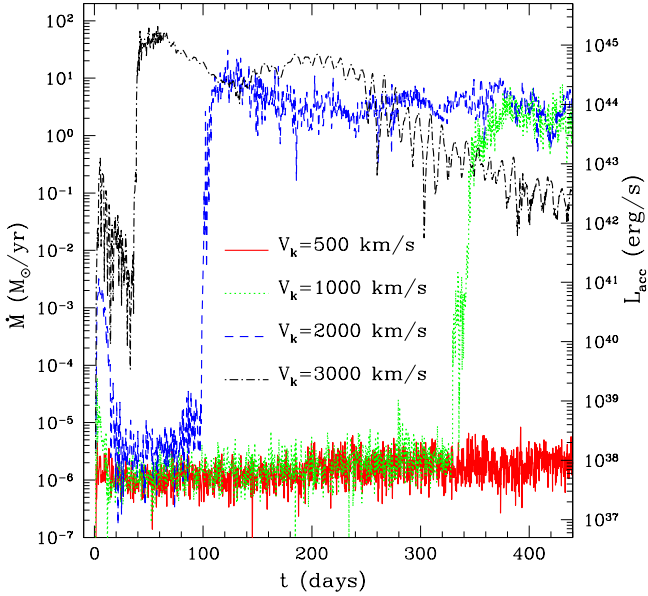}}
\vspace{-2.0cm}
\caption{Mass accretion rate at $r=r_{\rm{min}}$ for different values
  of the kick velocity in the large-size model
  \texttt{L.00}. 
}
\label{fig6}
\end{figure}

Interestingly, because the black hole is moving at very large
velocities in an ambient fluid which also has a non-negligible angular
momentum, a low-density region which resembles a horn-shaped
``cavity'' is produced in the downstream part of the flow when $V_{\rm
  k}= 3000 \,{\rm km/s}$ at $t=180\,{\rm d}$ after the merger. This
cavity is shown in Fig.~\ref{fig12} and its orientation is directly
related to the direction of the recoil velocity. A simple change of
sign in the recoil velocity, in fact, would rotate the cavity of 180
degrees around the black hole (not shown in Fig.~\ref{fig12}). The
formation of such a cavity is noticeable also in the simulations
of~\citet{Rossi2010}, where however it is not discussed. The cavity
leads to quasi-periodic variation of the accretion rate as clumps of
matter in the downstream of the flow enter the cavity and streams onto
the black hole (\cf the oscillations of ${\dot M}$ in Fig.~\ref{fig6}
after $t\sim 200\,{\rm d}$ for $V_{\rm K}=3000\,{\rm km/s}$). The
generation of this flow pattern has an interest of its own, being a
non-trivial variant of the Bondi-Hoyle accretion flow onto a moving
black hole and will be investigated with greater detail in a distinct
work.

The third and final difference is really a combination of the
phenomenology discussed above as it manifests itself in terms of the
accretion rate. This is shown in Fig.~\ref{fig6}, which reports the
mass accretion rate at $r=r_{\rm{min}}$ for different values of the
kick velocity in the large-size model \texttt{L.00}. As already
discussed with Fig.~\ref{fig5} for the small-size models, also in
these larger discs the accretion rate shows a rapid increase 
when the black hole ``meets'' the disc, reaching values $\dot M \simeq
10 M_\odot/{\rm yr}$, that are well above the Eddington one because of
the absence of the radiation-drag contribution. Differently from the
smaller discs, however, the lag between the merger and the increase of
the accretion rate is not linear but rather increases nonlinearly as
the recoil velocity is decreased. This is clearly shown in
Fig.~\ref{fig6}, where the accretion rate jumps to high values after
$\sim 35,{\rm d}$ for $V_{\rm k}=3000\,{\rm km/s}$, after $\sim
100,{\rm d}$ for $V_{\rm k}=2000\,{\rm km/s}$ and after almost one
year, \ie $\sim 330,{\rm d}$ for $V_{\rm k}=1000\,{\rm km/s}$. The
reason for this nonlinear response of the disc has to be found in the
fact that for smaller recoil velocities the time required by the black
hole to cross the initial cavity $\tau_{\rm cross}$ will be much
larger than the typical orbital revolution time. As an example, for
$V_{\rm k}=1000\,{\rm km/s}$ the timescale ratio is $\tau_{\rm
  cross}/\tau_{\rm c} \sim 33$, so that the disc will have sufficient
time to readjust itself to the new gravitational field and thus
redistribute its orbital angular momentum. In practice the inner edge
of the disc will move to larger radii as a result of the mass loss and
of the varied potential, thus delaying nonlinearly the contact with
the black hole and thus the steep increase in the accretion rate.

As a final remark we note that all of the phenomenology discussed here
for the model \texttt{L.00} has been found also for a large-size model
around a rapidly spinning black hole, namely \texttt{L.90}. Because
the overall differences are minute and of the order of few percent at
most, they will not be discussed here. The reasons behind these
similarities are rather obvious: the large-size discs have inner radii
that are too far from the black hole to be sensitive to the
spin-induced corrections which decay much more rapidly and as $1/r^3$.

\subsubsection{EM Luminosities}

Because of the relatively high temperature of the gas and of the
generation of a shock pattern, thermal bremsstrahlung is thought to be
an efficient emission mechanism through which circumbinary discs may
become visible in the electromagnetic spectrum \citep{Megevand2009,
  Corrales2009, Bode:2009mt, Anderson2009}. However,
thermal-bremsstrahlung emission from circumbinary is affected by a
serious problem which has been so far underestimated or not
sufficiently emphasized. This has to do with the fact that
bremsstrahlung cooling time is too short~\citep{Corrales2009} or,
stated differently, that the internal energy budget of the emitting
gas is not large enough to allow for the bremsstrahlung emission to
last but for a few seconds. This can be easily estimated as $t_{\rm
  {cool}}=E_{\rm int}/L_{{\rm BR}}$, with $E_{\rm int}$ and $L_{{\rm
    BR}}$ obtained from (\ref{int_enrgy}) and~(\ref{brem_geo}),
respectively.  For the large model \texttt{L.00} we have $E_{\rm
  int}\sim 3.4\times10^{50}{\rm erg}$ and $L_{{\rm BR}}\sim
2.8\times10^{49}{\rm erg/s}$ at time $t=0$ and this estimate remains
of the same order of magnitude during the evolution.  This yields to
$t_{\rm {cool}} \simeq 12\,{\rm s}$.  The situation is even worse if
we consider the transition to the relativistic regime.  In this case,
in fact, not only the bremsstrahlung emissivity is increased by a
factor\footnote{ It should be remarked, however, that when the
  electron become relativistic, \ie for $T\geq 5.9\times 10^9 K$,
  other emission mechanisms, such as inverse Compton or synchrotron
  (if a magnetic field is also present), are generally more
  efficient.} $\propto [1+4.4 T/(10^{10}
  K)]$~\citep{Rybicki_Lightman1986}, but also the collisions between
particles of the same species start contributing significantly to the
bremsstrahlung emission~\citep{Svensson1982} through radiation in
moments other than the electric dipole (which is strictly zero for
particles of the same species~\citep{Krolik1999}).

\begin{figure}
{\includegraphics[angle=0,width=9.0cm]{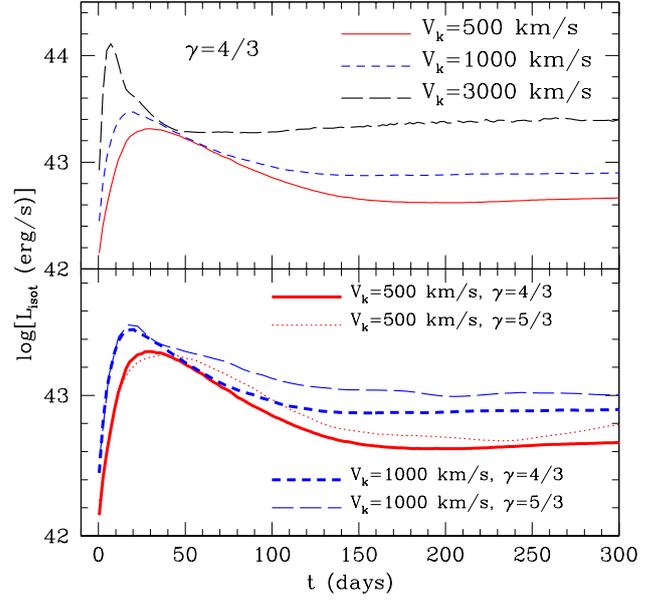}}
\vspace{-2.0cm}
\caption{Top panel: Luminosity computed in the isothermal evolution
  $L_{\rm isot}$ for different values of the kick velocity for model
  \texttt{L.00} and for a polytropic index $\gamma=4/3$.  Note the
  presence of a peak at about $\sim 20\,{\rm d}$ after the merger, of
  a binary with total mass $M\simeq 10^6 M_\odot$ and of a persistent
  luminosity for several days at values which are a factor of a few
  smaller. Bottom panel: Comparison of $L_{\rm isot}$ for model
  \texttt{L.00} when computed with a polytropic index $\gamma=4/3$
  (thick lines) or $\gamma=5/3$ (thin lines). The comparison is made
  for two reference recoil velocities and
  shows that the results are very similar, although a stiffer EOS
  leads to slightly larger luminosities.}
\label{fig10}
\end{figure}

Of course, there are also other factors that can work in favour of a
bremsstrahlung emission and which we have not taken into account. A
first one is that we have neglected the thermal bremsstrahlung
absorption, which is likely to enhance significantly the
bremsstrahlung cooling time by acting as a source of additional
internal energy. Moreover, it is also possible that the spiral shock
originating from the very central region dissipates considerably as it
propagates outwards, hence confining the bulk of the bremsstrahlung
luminosity from within a very small portion of the disc (we recall
that the bremsstrahlung luminosity is proportional to the volume
integral over the emitting source). Overall, while we cannot rule out
thermal bremsstrahlung as an emission mechanism from the circumbinary
disc, it is also evident to us that the luminosity estimates made so
far without a proper treatment of radiation transfer are excessively
optimistic.

A second possible estimate of the luminosity is given by the
accretion-powered luminosity, $L_{\rm acc}=\eta \dot M c^2$, where
$\eta$ is the radiative efficiency.  However, lacking any
  treatment of the radiation pressure effects, such an estimate would
  only provide misleading conclusions and therefore we will note use
  it hereafter.

A third and possibly more accurate estimate of the luminosity can be
made by assuming that all the changes in the temperature that are due
to a local compression will be dissipated as radiation. This idea,
proposed in Newtonian physics by~\citet{Corrales2009}, can be
summarized and implemented in a general relativistic context as
detailed below.

Consider the evolution of the disc with an equation of state
$p(T)=\rho k_b T/m_p$ and a specific internal energy given by
$\epsilon(T)=k_b T/[(\gamma-1) m_p] = \frac{3}{2}p/\rho$, where the
last equality has been obtained for $\gamma=5/3$. In general, there is
no necessity to evolve the energy equation in an isothermal evolution
since the energy can be computed directly from the temperature and the
latter is constant by construction. However, the internal energy can
be nevertheless evolved in time with the only aim of computing the
difference $\rho[\epsilon - \epsilon(T)]$, which is then assumed to be
radiated instantaneously. The relativistic equation for the evolution
of the total internal energy density $e\equiv\rho(1+\epsilon)$
is~\citep{Anile_book}
\be
\label{internal_energy}
u^\mu\nabla_\mu e + (e+p)\Theta=0,
\ee
where $\Theta\equiv\nabla_\mu u^\mu$ is the expansion of the
fluid. The continuity equation $\nabla_\mu(\rho u^\mu)=0$ can then be
used to rewrite Eq.~(\ref{internal_energy}) as
\bea
\label{internal_energy1}
\partial_t(\sqrt{\gamma}W\rho\epsilon)+\partial_i[\sqrt{\gamma}\rho\epsilon
  W(\alpha
  v^i-\beta^i)] = && \nonumber \\ 
&& \hskip -2.5cm -p\partial_t(\sqrt{\gamma}W) -p\partial_i(\alpha\sqrt{\gamma}u^i) 
\ .
\eea

One aspect to note is that Eq.~(\ref{internal_energy1}) is not written
in a conservative form because of the derivatives on the right hand
side acting on the flow variables. While this is not ideal within our
formulation of the hydrodynamics equations, the modifications are
minimal. Indeed, since the Lorentz factor does not change
significantly during the evolution, it is reasonable to neglect the
term $\propto \partial_t(\sqrt{\gamma}W)$, while the spatial
derivatives of the term $\partial_i(\alpha\sqrt{\gamma}u^i)$ can be
treated with standard finite-difference methods without a significant
loss of accuracy across the discontinuities.

In practice we have assumed an initial temperature for the disc which
is uniform in space and set it to be $T_0=\frac{1}{2} T_{\rm c}$,
where $T_{\rm c}$ is the maximum temperature at the center of the disc
(\cf Table~\ref{tab1}). An estimate of the luminosity is then
trivially computed by performing at each timestep a volume integration
of the difference $\rho[\epsilon - \epsilon(T)]$ and by dividing it by
the simulation timestep. Finally, we reset the specific internal
energy to $\epsilon = \epsilon(T_0)$, so as to guarantee that the
evolution is effectively isothermal.

\begin{figure*}
{\includegraphics[angle=0,width=9.0cm]{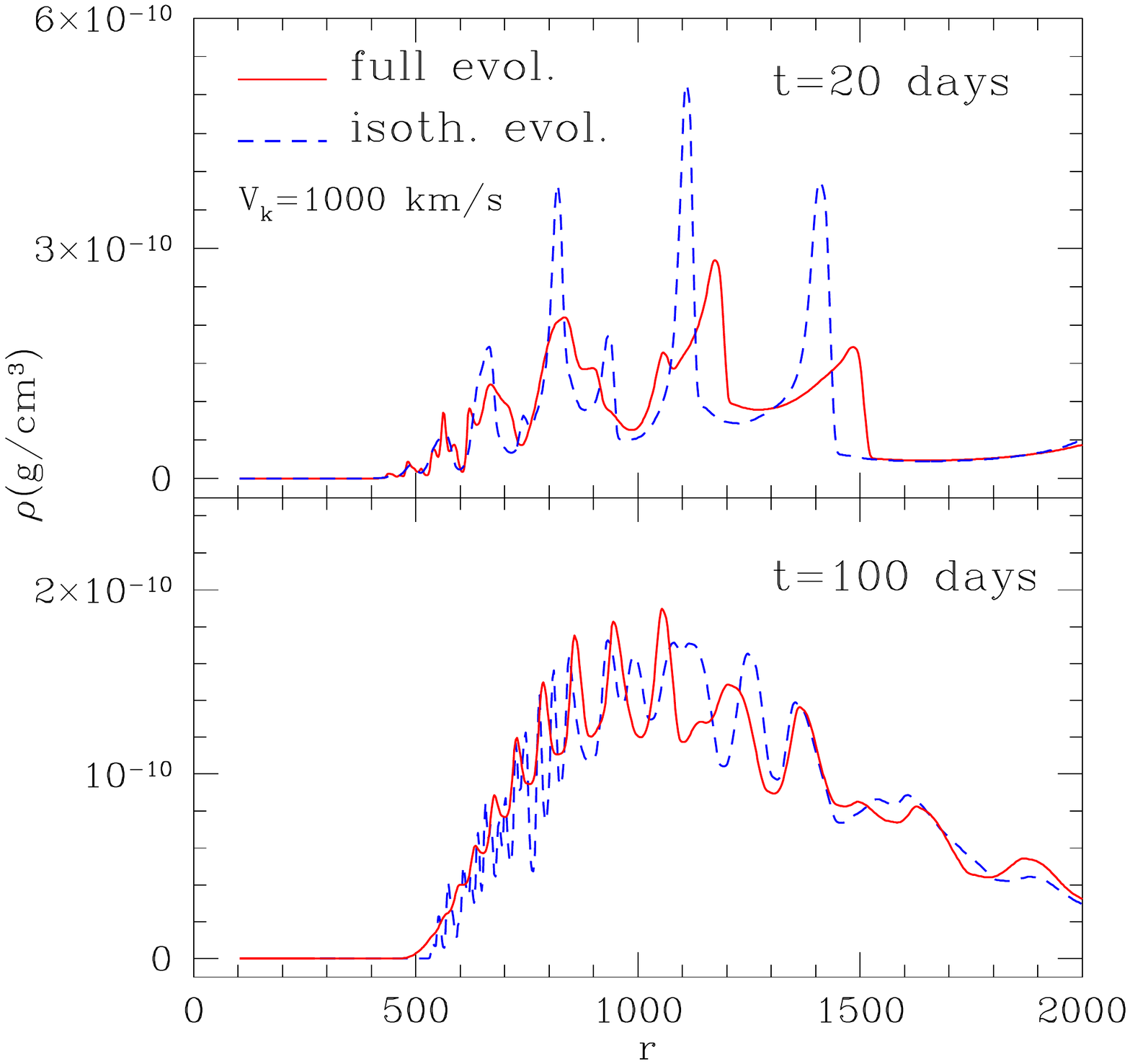}}
\hskip 0.5cm
{\includegraphics[angle=0,width=9.0cm]{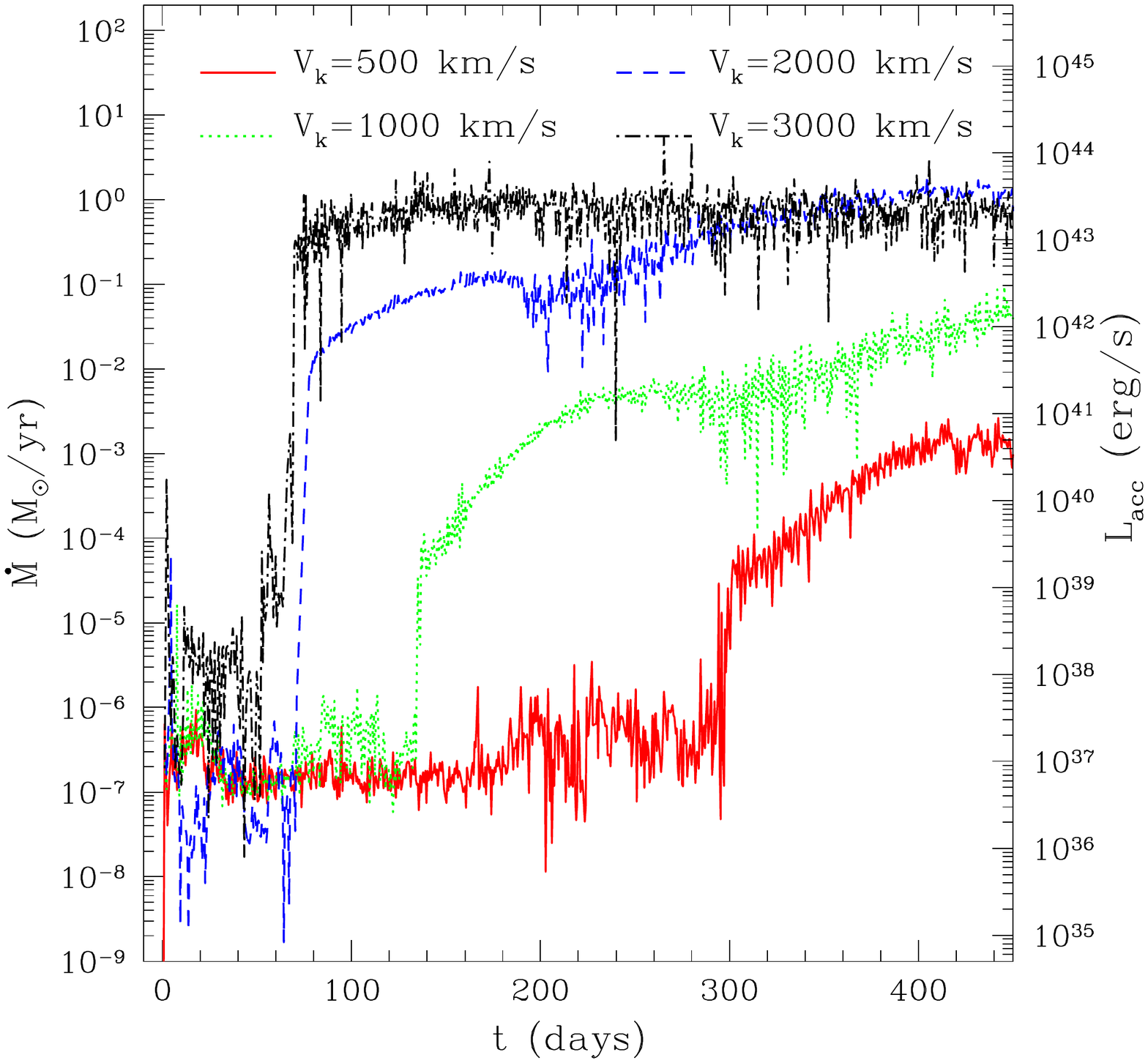}}
\vspace{-2.0cm}
\caption{Left panel: Comparison of the rest-mass density along the
  $\phi=0$ direction as obtained with the standard evolution of the
  energy equation (red solid line) and with the isothermal evolution
  (blue dashed line), at two different times as shown in the two
  panels. The data refers to model \texttt{L.00} with a recoil of
  $1000\,{\rm km/s}$. Right panel: Mass accretion rate at
  $r=r_{\rm{min}}$ for the isothermal evolution and different values
  of the kick velocity in the large-size model \texttt{L.00}. This
  panel should be compared with Fig.~\ref{fig6} and which refers to an
  evolution of the energy equation.}
\label{fig9}
\end{figure*}

The top panel of Fig.~\ref{fig10} shows the luminosity computed in
this way for model \texttt{L.00} with a polytropic index $\gamma=4/3$
and for different values of the recoil
velocity. Following~\citet{Corrales2009}, the values reported have
been computed by neglecting the negative contributions to the
luminosity that are produced in regions experiencing
rarefactions. Because when accounted for these negative contributions
typically yield values that are one order of magnitude smaller, the
values in Fig.~\ref{fig10} should be taken as upper limits to the
emitted luminosity~\citep[see][]{Corrales2009}. Clearly, the
evolution of the emitted energy has a peak that is larger for stronger
recoils and that appears at $t\sim 33, 18$, and $7$ d after the merger
for $V_{\rm k}=300, 1000$ and $3000 \,{\rm km/s}$, respectively. While
the peaks are the consequence of the strong shocks that are produced
in the inner parts of the disc as the latter approaches the black
hole, the asymptotic values of the isothermal luminosity are instead
produced by the local compressions in the disc. As such, the peaks in
the luminosity are not related to the encounter of the black hole with
the disc and therefore they are not correlated with the increase in
the mass accretion rate, which in general takes place at later times
(\cf right panel of Fig.~\ref{fig9}).  The bottom panel of
Fig.~\ref{fig10}, on the other hand, shows a comparison of $L_{\rm
  isot}$ for model \texttt{L.00} when computed with a polytropic index
$\gamma=4/3$ (thick lines) or $\gamma=5/3$ (thin lines). The
comparison is made for two reference recoil velocities of $V_{\rm
  k}=300\,{\rm km/s}$ and $V_{\rm k}=1000\,{\rm km/s}$
 and it shows that the results are very similar,
although a stiffer EOS leads to slightly larger luminosities.

Overall, our estimates of the luminosity computed within the
isothermal evolution approximation confirm those
by~\citet{Corrales2009} even though the temperature of our large size
model is one order of magnitude larger than that reported in Fig.~7
of~\citet{Corrales2009}. While we believe that these estimates of the
luminosity are the most reasonable ones that can be obtained with a
code that is intrinsically unable to account for radiation losses, we
should also stress that the isothermal evolution by itself provides a
less reliable description of the overall dynamics. This is evident,
for example, from Fig.~\ref{fig9}, whose left panel offers a
comparison of the rest-mass density profiles along $\phi=0$ as
obtained with the standard evolution of the energy equation (red solid
line) and with the isothermal evolution (blue dashed line) for the
large-size model \texttt{L.00} and a recoil of $V_{\rm k}=1000\,{\rm
  km/s}$. Note that the isothermal evolution tends to increase the
density gradients, especially during the initial phases, which is also
when the peak in the luminosity appears in the light curves. Also
shown in the right panel of Fig.~\ref{fig9} is the mass accretion rate
at $r=r_{\rm{min}}$ for the isothermal evolution and different values
of the kick velocity. This panel shows that since the discs are more
sensitives to compressions, their response to the recoil is different
and, in particular, takes place earlier than in the full-evolution
case (\cf with Fig.~\ref{fig6}). Furthermore, with the exception of
the very large kick, the mass accretion rate does not show a very
large jump, but rather a first small jump followed by smooth increase
to the final asymptotic behaviour which is reached at times comparable
to those of the full evolution. This is clearly the case for $V_{\rm
  k}=1000\,{\rm km/s}$ and is due to the fact that accretion starts
earlier for this matter whose energy has been decreased by the
radiative losses.

In conclusion, and in spite of the caveats made above, we believe that
luminosities as large as few $L \simeq 10^{43} \ {\rm erg/s} $ should
be expected at about $\sim 30\,{\rm d}$ after the merger of a binary
with total mass $M\simeq 10^6 M_\odot$ and that these luminosities
should persist for several days to values which are a factor of a few
smaller.

\section{Conclusions}
\label{Conclusions}

We have presented the results of two-dimensional general relativistic
numerical simulations of small \textit{and} extended circumbinary
discs in the post-merger phase of the merger, when the disc reacts to
the mass loss of the central black hole and to its recoil
velocity. 
Our analysis benefitted from being able to capture accurately the
dynamics of the perturbed disc in the relativistic regime, thus
allowing us to investigate the dependence of the accretion rate on the
black-hole spin and on the kick velocity. Furthermore, by considering
discs that are quasi-Keplerian, extended and at a large distance from
the binary, we were able to consider realistic scenarios
even in the general relativistic framework.

Another important aspect of our work is the use of a novel and
accurate technique to construct a ``shock detector'' and hence to
determine where, within the flow, the shocks produced by the recoil
are located. This, in turn, has allowed us to assess under what
conditions a spiral-shock pattern can develop, produce a variability
in the accretion rate and, hence, in the luminosity. Our relativistic
shock detector (for which we also present a Newtonian equivalent in
Appendix~\ref{appendixA}) is based on the analysis of the initial
states of the Riemann problem solved at each cell interface and can
therefore determine the location of the shock with the same resolution
as that of the spatial grid, revealing that the previously proposed
criteria for the occurrence of the shock are often inaccurate.

Overall, we can confirm within a general relativistic regime many of
the results found previously in Newtonian or pseudo-Newtonian
gravity. More specifically, we find that for discs that are
sufficiently small and close to the black-hole, a regular spiral-shock
develops as a result of the recoiling black hole. The strength, shape
and persistence of the shocks, however, depend sensitively on both the
size of the tori and on the intensity of the recoil. As a result,
while the spiral shock is stable over many orbital periods in the case
of small discs subject to small recoils, it never develops or is
rapidly destroyed in discs that are large and subject to large recoil
velocities. It is worth noting that the typical velocity jumps at the
shocks are $\Delta v \lesssim 2.5\times 10^{-4}$, even with a kick
velocity $V_{\rm k}=3000 \ {\rm km}/{\rm s}$. It is therefore possible
that such shocks may be damped by dissipative viscous processes or
radiative losses.

Besides the interesting shock properties described above, we also
found that the disc dynamics is only very weakly dependent on the
black-hole spin. The latter influences only the small-size tori by
modifying the accretion rate and by leading to smaller accreted masses
per unit time for more rapidly spinning black holes. Finally, we found
that even in the limit of a vanishing kick velocity and as long as a
mass loss is present, the disc goes through a phase of regular
oscillations characterized by the excitation of the $p$ modes of the
disc; this is then followed by the appearance of a spiral-shock
pattern generated by non-axisymmetric instabilities affecting the
disc. This opens the door to the possibility that quasi-periodic
oscillations are observed in the post-merger phase of SMBBH.

Computing the EM counterpart to the merger event represents of course
a fundamental aspect of our investigation. In contrast with other
works, however, we have questioned the estimates of the bremsstrahlung
luminosity when computed without properly taking into account the
radiation transfer. The energetic reservoir available for the EM
emission is, in fact, too small to yield but unrealistically short
cooling times. While we cannot rule out thermal bremsstrahlung as the
main EM emission mechanism, we believe that the
bremsstrahlung-luminosity estimates made so far without a proper
treatment of radiation transfer are excessively optimistic. A somewhat
realistic estimates of the emitted luminosity can be obtained by
assuming that the internal-energy enhancements due to local
compressions in the perturbed disc are immediately radiated away, \ie
by considering an ``isothermal'' evolution like the one recently
investigated in Newtonian physics by~\citet{Corrales2009}. In this
case, and despite the fact that the isothermal evolution tends enhance
the compressibility of the fluid, we find that the energy emitted can
reach a peak value above $L \simeq 10^{43} \ {\rm erg/s} $ at about
$\sim 30\,{\rm d}$ after the merger of a binary with total mass
$M\simeq 10^6 M_\odot$ and persist for several days at values which
are a factor of a few smaller. If confirmed by more sophisticated
calculations such a signal could represent an interesting EM
counterpart of the merger of binary black-hole system.

As a final remark we note that while a rather robust picture is
emerging from the collective work done so far on the post-merger
dynamics of the circumbinary disc around a SMBBH, much remains to be
done to compute realistically the resulting EM emission. Important
improvements to the treatment considered here must include the
presence of a magnetic field, a proper treatment of the radiation
transfer and the extension to three-dimensional calculations. All of
these will be the focus of our future work on this subject.


\begin{acknowledgements}
We are grateful to Marek Abramowicz and Constanze Roedig for useful
discussion. The computations were performed on the Damiana cluster at
the AEI and on the IBM/SP6 of CINECA (Italy) through the
``INAF-CINECA'' agreement 2008-2010. This work was supported in part
by the DFG grant SFB/Transregio~7.
\end{acknowledgements}


\appendix
\section[]{Newtonian version of the shock detector}
\label{appendixA}

In this Appendix we provide the basic expressions that allow to build
the Newtonian version of the relativistic shock detector described in
Sec.~\ref{Shock_detection}. The logic, of course, is exactly the same
and it consists of the following steps.

\begin{enumerate}

\item
First choose the direction, say the $x-$direction, along which to
monitor the generation of shock waves.
\item
Given any two adjacent cells, label with $1$ the cell with higher
pressure and with $2$ the other one.
\item
Compute the relative velocity $v_{12}=v_1-v_2$ along the $x-$direction
and compare it with the value 
~\citep[see][\textsection 100]{Landau-Lifshitz6}
\begin{equation}
({\widetilde v}_{12})_{_{SR}}\bigg\vert_{\rm Newt} =
	-\frac{2}{\gamma-1}c_s(p_1)
	\left[1-\left(\frac{p_2}{p_1}
	\right)^{(\gamma-1)/2\gamma}\right] \ ,
\end{equation}
where $\gamma$ is the adiabatic index of the gas and $c_s(p_1)$ is the
sound speed in the state $1$. It is interesting to note that in the
Newtonian case the tangential velocities do not affect the actual
value of the threshold $({\widetilde v}_{12})_{_{SR}}$, which,
therefore, maintains the same expression as for one-dimensional
flows. Indeed, the fact that in the relativistic regime the threshold
given by (\ref{capo2_analytic}) does depend on the tangential
velocities is at the origin of the appearance of new relativistic
effects discussed in~\citet{Rezzolla02}.
\item
A shock is therefore detected if
\be
\label{condition_newtonian}
v_{12}>({\widetilde v}_{12})_{_{SR}}\bigg\vert_{\rm Newt} .
\ee
\end{enumerate}

The procedure is repeated for as many directions as the dimensions of
the problem. As commented in the main text, we note here too that it
may be necessary at times to filter-out the smallest shocks and
therefore make the condition for shock detection more stringent. In
analogy with Eq.~(\ref{condition2}), the following condition can then
be implemented
\begin{equation}
v_{12}>{\tilde v}_{12}=({\tilde v}_{12})_{_{SR}}\bigg\vert_{\rm
  Newt} +\chi\left[({\tilde v}_{12})_{_{2S}}\bigg\vert_{\rm Newt} - 
  ({\tilde v}_{12})_{_{SR}}\bigg\vert_{\rm Newt}\right], 
\end{equation}
where~\citep[see][\textsection 100]{Landau-Lifshitz6}
\begin{equation}
({\tilde v}_{12})_{_{2S}}\bigg\vert_{\rm Newt} =
(p_1 - p_2) \sqrt{\frac{2}{\rho_2\left[
(\gamma + 1)p_1 + (\gamma - 1)p_2 \right]}}, 
\end{equation}
and where the parameter $\chi \in [0,1]$ needs to be tuned to the
desired level of accuracy.

\bibliographystyle{aa}
\bibliography{biblio/aeireferences}

\end{document}